\documentclass[a4paper,5pt,intlimits,twopages]{iopart}

\usepackage{xcolor}
\usepackage[colorlinks=true,
linkcolor=blue,
citecolor=red,
urlcolor=teal]{hyperref}

\expandafter\let\csname equation*\endcsname\relax
\expandafter\let\csname endequation*\endcsname\relax

\usepackage[margin=2cm]{geometry}
\usepackage[utf8]{inputenc}
\usepackage[tbtags]{amsmath}
\usepackage{amsthm,amssymb,dsfont,mathrsfs,bbm,xfrac,tabularx,booktabs}
\usepackage{comment}
\usepackage{mathtools}
\usepackage{stmaryrd}
\usepackage{scalerel}
\usepackage{xcolor}
\usepackage{subcaption}
\usepackage{xfrac}

\usepackage{hyperref}
\usepackage[capitalise]{cleveref}
\usepackage{tikz}

\usepackage{lmodern}
\usepackage[T1]{fontenc}

\allowdisplaybreaks

\begin{document}

\title{Overlap Gap and Computational Thresholds in the Square Wave Perceptron}


\author{Marco Benedetti$^{1}$, Andrej Bogdanov$^{2}$, Enrico M. Malatesta$^{1}$, Marc Mézard$^{1}$, Gianmarco Perrupato$^1$, Alon Rosen$^1$, Nikolaj I. Schwartzbach, Riccardo Zecchina$^1$}

\address{$^1$Department of Computing Sciences and Bocconi Institute for Data Science and
	Analytics (BIDSA), Bocconi University, 20136 Milano, Italy}
\address{$^2$University of Ottawa, Canada}

\begin{abstract}
	Square Wave Perceptrons (SWPs) form a class of neural network models with oscillating activation function that exhibit intriguing ``hardness'' properties in the high-dimensional limit at a fixed constraint density $\alpha = O(1)$. In this work, we examine two key aspects of these models. 
	
	The first is related to the so-called \emph{overlap-gap property}, that is a disconnectivity feature of the geometry of the solution space of combinatorial optimization problems proven to cause the failure of a large family of solvers, and conjectured to be a symptom of algorithmic hardness. We identify, both in the storage and in the teacher-student settings, the emergence of an overlap gap at a threshold $\alpha_{\mathrm{OGP}}(\delta)$, which can be made arbitrarily small by suitably increasing the frequency of oscillations $1/\delta$ of the activation. This suggests that in this small-$\delta$ regime, typical instances of the problem are hard to solve even for small values of $\alpha$.  
	
	Second, in the teacher-student setup, we show that the recovery threshold of the planted signal for message-passing algorithms can be made arbitrarily large by reducing $\delta$.
	
	These properties make SWPs both a challenging benchmark for algorithms and an interesting candidate for cryptographic applications.
\end{abstract}


\section{Introduction}

Perceptrons are families of functions introduced in the 1950s to mathematically model the key ability of intelligence to make associations \cite{rosenblatt1958perceptron}. More specifically, a perceptron can be defined in terms of an ``activation function'' $\varphi(\bullet):\mathds{R}^N\rightarrow \mathcal{B}_{out}\subseteq\mathds{R}$, and a vector of ``synaptic'' weights $\boldsymbol{w}\in \mathcal{B}_{s}^{N}\subseteq\mathds{R}^N$, and it associates an information, encoded in a input pattern $\boldsymbol{x}\in\mathds{R}^N$, with a label $y\in\mathcal{B}_{out}$, through the map $y=\varphi\big(\boldsymbol{w}\cdot \boldsymbol{x}/\sqrt{N}\big)$. 
In our discussion, we focus on \emph{binary perceptrons}, where the alphabet of synapses and labels, respectively, $\mathcal{B}_{s}$ and $\mathcal{B}_{out}$ are binary, i.e.\ $\mathcal{B}_{out}=\mathcal{B}_{s}=\{-1,+1\}$. 

\begin{figure}
	\centering
	\includegraphics[width=0.3\linewidth]{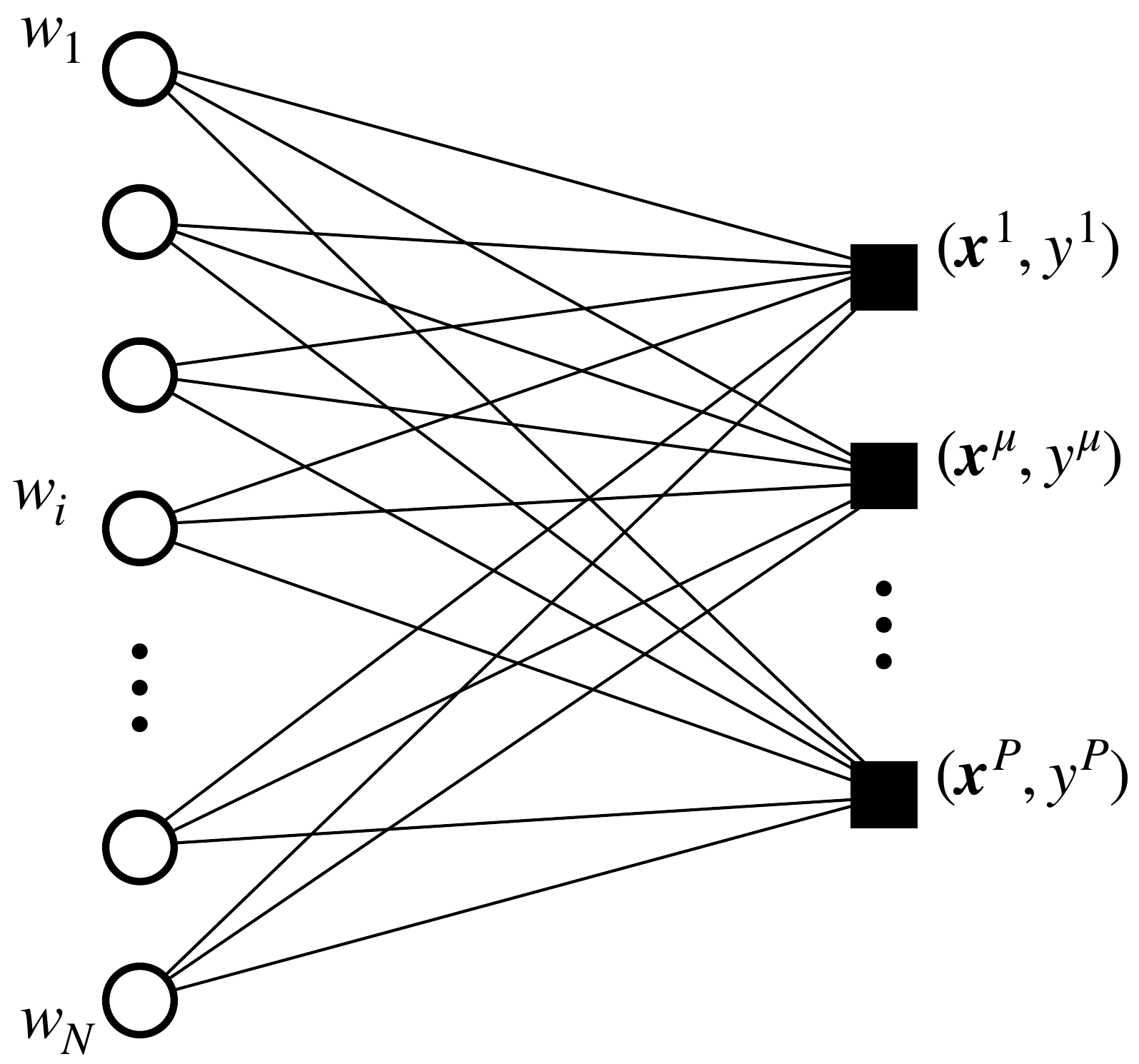}
	\caption{Factor graph representing the constraint satisfaction problem defined by \eqref{eq:conditions}. Circles represent the variables, that in this case are the weights $w_i$, and squares represent the constraints, given by the input-output associations $(\boldsymbol{x}^{\mu},y^{\mu})$.}
	\label{fig:factor-graph-perceptron}
\end{figure}

Perceptron-based learning prototypes have been extensively studied by the statistical physics community, showing that despite their simplicity, which enables analytical investigations, these models exhibit complex behaviors, such as distinct thermodynamic phases separated by phase transitions. Two fundamental learning prototypes are the so-called ``storage'' and ``teacher-student'' problems. In both cases, the objective is to find a vector of weights $\boldsymbol{w}\in\{-1,+1\}^{N}$ s.t.\ the perceptron fits a database of $P$ given associations $\{(\boldsymbol{x}^{\mu},y^{\mu})\}_{\mu=1}^P\equiv \mathcal{D}$. In the statistical physics approach, this database is the source of ``quenched disorder''. In other words, one requires that the following constraints are satisfied by $\boldsymbol{w}$ (see also \cref{fig:factor-graph-perceptron}):
\begin{equation}
	\label{eq:conditions}
	y^{\mu}=\varphi\left(\frac{1}{\sqrt{N}}\boldsymbol w\cdot\boldsymbol x^{\mu}\right) \quad \forall\,\, \mu\in[P]\,.
\end{equation}
The associations are chosen stochastically. In particular, the inputs $\boldsymbol{x}^{\mu}\in \mathds{R}^N$ are i.i.d.\ vectors with i.i.d.\ elements $x_i^{\mu}\sim\mathcal{N}(0,1)$, $i\in[N]$. The difference between the storage and the teacher-student settings lies in the generation procedure for the labels:
\begin{itemize}
	\item in the \emph{storage} setting the $y^\mu$'s are i.i.d.\ Rademacher: $y^\mu=\pm 1$ with probability $1/2$;
	\item in the ``teacher-student'' (or \emph{planted}) setting one draws a ``teacher'' set of weights, $\boldsymbol{w}^*\in\{-1,1\}^N$, with i.i.d.\ Rademacher elements  $w^*_i$, $i\in[N]$, and the labels are given by:
	\begin{equation}
		y^{\mu}=\varphi\left(\frac{1}{\sqrt{N}}\boldsymbol {w}^*\cdot\boldsymbol x^{\mu}\right)\quad \forall\, \mu\in[P]\,.
	\end{equation}
\end{itemize}
Different activation functions have been studied in the literature. A notable case is that of the asymmetric binary perceptron~\cite{Gardner_1989} (ABP), defined by the activation $\varphi(\bullet)=\text{sgn}(\bullet)$, where \eqref{eq:conditions} corresponds to an integer program version of a binary linear classification task. Another well-known case is that of the symmetric binary perceptron~\cite{aubin2019storage} ($\text{SBP}_{\kappa}$), whose storage version can be defined by setting all the $y^\mu$'s equal to one and taking the activation as $\varphi(\bullet)=\text{sgn}(\kappa-|\bullet|)$, where $\kappa$ is a parameter. In this case, \eqref{eq:conditions} is a minimization problem of the scalar products between the patterns and $\boldsymbol{w}$, and has connections with combinatorial discrepancy theory \cite{spencer1985six,matousek2009geometric,gamarnik2022algorithms} and number partitioning \cite{vafa2025symmetric}. 

Interestingly, the previous perceptron problems are not only studied as toy models for learning, but are also considered insightful examples of constraint satisfaction problems (CSPs) by various communities interested in aspects of algorithmic hardness. Examples range from the statistical physics of glasses \cite{franz2016simplest} to theoretical computer science \cite{PX21,ALS22}. In fact, depending on the fraction of the number of patterns $\alpha=P/N$, which plays the role of a control parameter, the behavior of polynomial-time algorithms designed to find a solution to constraints \eqref{eq:conditions} could change dramatically. In particular, in the limit $P,N\rightarrow \infty$ held at constant ratio $\alpha$, there are regions of $\alpha$ where the constraints \eqref{eq:conditions} are satisfiable w.h.p.\, but no polynomial-time algorithm is known to find solutions. For the teacher-student problem, in cryptography parlance, the network is a \emph{one-way function} (OWF) for these parameters. A fundamental open question concerns the mathematical characterization of these ``hard'' regions. From statistical physics investigations, the idea emerged that hardness in CSPs can be understood in terms of changes in the geometrical properties of their solution space \cite{mezard2002analytic,mezard2002random,mezard2005clustering,achlioptas2006solution,monasson1999determining}. 
A transition between a regime where the densest cores of solutions are immersed in a huge connected structure, and a regime where this structure breaks up is conjectured to mark the onset of algorithmic hardness. This transition, referred to as the \emph{local entropy} transition, was found in the solution space of a binary perceptron problem \cite{baldassi2015subdominant}, matching the threshold where the best known polynomial solvers for the problem stop working. Building upon this geometric point of view, \cite{gamarnik2019landscape} defined a disconnectivity feature of the space of solution of CSPs termed \emph{overlap gap property} (OGP). Beyond capturing the topological features of the solution space emerging at the local entropy transition, the OGP — or stronger variants such as the ensemble-OGP — provably imply the failure of broad classes of efficient algorithms \cite{gamarnik2021overlap}. The observation that the best available efficient solvers for certain hard CSPs are hindered by the presence of an OGP has led to the conjecture that the OGP serves as a general predictor of algorithmic hardness across a wide range of problems, whose precise characterization remains an open question\footnote{There exist problems where OGP alone is not predictive of hardness (see~\cite{InferenceHighDimensionalLinear,songCryptographicHardnessLearning2021,liEasyOptimizationProblems2025}).}.\\
In this paper, we study the onset of OGP and the consequent conjectured hardness thresholds for the \textit{teacher-student} and \textit{storage} problems in a family of oscillating activation functions of the form
\begin{equation}
	\label{eq:defSWP}
	\varphi_{\delta}(h)=-\text{sgn}\left(\sin{\left(\frac{\pi}{\delta}\,h\right)}\right)\,,
\end{equation}
that we call \emph{square wave perceptrons} (SWPs). We show that SWPs exhibit extreme hardness properties in the limit of small $\delta$. We identify, both in the storage and in the teacher-student settings, the emergence of an overlap gap at a threshold, which can be made arbitrarily small by suitably increasing the frequency of oscillations $1/\delta$ of the activation function in \cref{eq:defSWP}. This suggests that in this small-$\delta$ regime, typical instances of the problem are hard to solve even for small values of $\alpha$.
In the teacher-student setup, we also show that the recovery threshold of the planted signal for message-passing algorithms can be made arbitrarily large by reducing $\delta$, similarly to certain adversarially planted random boolean satisfiability problems \cite{feldman2015complexity}. In a recent line of research, the success of efficient algorithms has been linked to the existence of regions of densely packed solutions (see for example \cite{baldassi2019}). The intuition behind the hardness of the SWP as $\delta\to0$, is that increasing the frequency of oscillations should destroy such dense entropic regions, as the labels become more and more sensitive to the weights.\\
These properties make SWPs both a challenging benchmark for algorithms and an interesting candidate for cryptographic applications. 
In particular, in \cite{benedetti2025neural} a design for Collision Resistant Hash Functions has been proposed, leveraging some aspects of computational hardness in the SWP. In short, Collision Resistant Hash Functions are functions for which finding two distinct inputs with the same output is computationally hard. As Collision Resistant Hash Functions are of fundamental importance, being at the basis of multiple cryptographic protocols and security guarantees \cite{goldwasser2019knowledge,blum1983coin,diffie2022new,yao1982protocols}, it is interesting to explore other nuances of computational hardness stemming from SWPs activation in more canonical settings, as the \textit{storage} and \textit{teacher-student} problem.\\

The rest of the paper is organized as follows. In \cref{ssec:IntroOGP}, we discuss the thermodynamic properties of perceptron problems, and we introduce the notion of ``overlap gap'' property. We also discuss the analytic framework for the study of the OGP in perceptron problems with generic activation functions. In \cref{sec:Storage}, we study the storage problem for the SWP. We start by computing the satisfiability threshold and then move to the analytical determination of the $m$-OGP thresholds within an annealed and an RS ansatz for different values of $m$ and $\delta$. We also study the limit $\delta\rightarrow 0$: in this regime, we find that solutions exist up to $\alpha=1$, and all the region $0<\alpha<1$ presents an overlap gapped phase. In \cref{sec:teacherstudent} we study the teacher-student problem for the SWP. We compute the threshold $\alpha_T(\delta)$ above which the planted solution becomes the unique solution to the problem. Similarly to the storage case, we compute the $m$-OGP thresholds within an annealed and an RS ansatz for different values of $m$ and $\delta$ and analyze the limit $\delta\rightarrow 0$. Finally, we analyze the algorithmic problem of recovering the planted signal above $\alpha_T(\delta)$. 

\section{Phase Transitions and the Overlap Gap Property}
\label{ssec:IntroOGP}

By taking the limit $N,P\rightarrow \infty$, at fixed constraint density $P/N\equiv\alpha$, the uniform measure on the space of solutions of typical instances of the perceptron problem undergoes structural changes as a function of $\alpha$. These ``phase transitions'' are not peculiar of perceptrons, and have been the subject of extensive studies by the statistical physics community in a variety of CSPs~\cite{mezard2009information, kirkpatrickselman1994,mezard2002analytic,krzakala2007gibbs}. The storage problem is characterized by a \emph{satisfiability} threshold $\alpha_c(\varphi)$, which depends on the activation $\varphi(\bullet)$, s.t.\ for $\alpha>\alpha_c(\varphi)$, typical instances do not admit solutions with high probability. Interestingly, both ABP and $\text{SBP}_{\kappa}$ display statistical-to-computational gaps: despite the fact that below $\alpha_c(\varphi)$ finding a solution is information-theoretically feasible, the best-known algorithms running in polynomial time stop working at a value of $\alpha$ \emph{strictly} below the satisfiability threshold. For example, in $\text{SBP}_{\kappa}$, the satisfiability threshold behaves like $\alpha_c(\kappa)\approx (\log_2{1/\kappa})^{-1}$ \cite{aubin2019storage}, in the regime where $\kappa\rightarrow 0$. This is in stark contrast with the behavior of the best known algorithmic guarantee for finding a solution, the Bansal and Spencer \cite{bansal2020line} algorithm, which works for $\alpha=O(\kappa^2)$ in the limit $\kappa\rightarrow 0$. Therefore, the storage capacity is asymptotically much larger than the algorithmic threshold. Analogous statistical-to-computational gaps are also found in the planted setting and in the ABP case~\cite{baldassi2015subdominant,Baldassi2015,braunstein2006learning,baldassi2021unveiling}. Remarkably, in some particular cases, the existence of these gaps in average case CSPs has been linked with worst-case complexity theory via worst-case to average-case reductions. In particular, it was recently shown that the average-case hardness of $\text{SBP}_{\kappa}$, and another variant of the problem, can be based on worst-case hardness assumptions on approximate shortest vector problems on lattices \cite{vafa2025symmetric}. 

Since the seminal works on random boolean satisfiability, the origin of these gaps has been started to be understood in relation to the geometric properties of the solution space of CSPs \cite{mezard2002analytic,mezard2002random}. In the specific case of perceptrons, for any $\alpha>0$, typical solutions according to the uniform measure are \emph{singletons} separated by $O(N)$ spin flips (golf-course scenario)~\cite{KrauthMezard1989,huang2014origin}. Finding such ``isolated'' solutions is conjectured to be algorithmically hard \cite{mezard2005clustering,achlioptas2006solution,monasson1999determining}. Indeed, solutions returned by polynomial-time algorithms working for $\alpha=O(1)$ are found to belong to connected ``entropic'' clusters that are atypical with respect to the uniform measure. In \cite{baldassi2015subdominant,baldassi2021unveiling}, the thermodynamics of these entropic states is studied with statistical physics techniques in the case of the ABP. Their analysis identifies a transition point $\alpha_U$ in the structure of the ``entropic'' clusters, that is interpreted as a separation between a regime ($\alpha<\alpha_U$) where the densest cores of solutions are immersed in a huge connected structure, and a regime ($\alpha_U<\alpha$) where this structure breaks up. The authors conjecture that this transition is related with the onset of algorithmic hardness, in agreement with numerical experiments. This same transition has been studied and analytically determined also in other CSPs such as spherical weight perceptrons and one hidden layer neural networks which store random patterns~\cite{baldassi2022learning,baldassi2023typ,baldassi2019}. 


Building upon the idea of a connection between the clusterization properties of the space of solutions and the onset of algorithmic hardness, it has been shown that the presence of a specific disconnectivity property called overlap-gap property (OGP), can be rigorously proven to imply the failure of stable algorithms \cite{gamarnik2021overlap,gamarnik2025turing}. Informally, thinking of an algorithm $\mathcal{A}$ as a map from the instance space to the solution space of a CSP, stability of $\mathcal{A}$ requires that a small perturbation of the input results in a small perturbation of the output. A large class of algorithms has been shown to satisfy this stability property in different CSPs, e.g.\ the Kim-Roche algorithm for the $\text{SBP}_{\kappa}$ \cite{gamarnik2022algorithms}, approximate message passing (AMP) \cite{Gamarnik_2021}, low degree polynomials \cite{Gamarnik_2020,Wein_2022}, Langevin dynamics \cite{Gamarnik_2020} and low depth Boolean circuits \cite{Gamarnik_2024}.

There are several variants of OGP. We refer the reader to \cite{gamarnik2021overlap} for references and details. For the purposes of this work, we define a version of the so-called $m$-OGP that we are going to use in the next sections. First of all, given an instance $\mathcal{D}=\{(\boldsymbol{x}^{\mu},y^{\mu})\}_{\mu=1}^P$ of the disorder, we define the set $\mathcal{S}_m(q,\epsilon,\mathcal{D})$ of all sets of $m$ configurations $\boldsymbol{w}_1,\dots,\boldsymbol{w}_m$, s.t.\ the following two properties hold:
\begin{enumerate}
	\item for each $a\in[m]$, $\boldsymbol{w}_a$ satisfies the constraints \eqref{eq:conditions}.
	\item for each $a\neq b\in[m]$, 
	\begin{equation}
		q < \frac{\boldsymbol{w}_a\cdot\boldsymbol{w}_b}{N} < q+\epsilon\,.
	\end{equation}
\end{enumerate}
We say that the $m-$OGP property holds if there exist $q$ and $\epsilon$ such that:
\begin{equation}
	\text{Pr}_{\mathcal{D}}\left[\mathcal{S}_m(q,\epsilon,\mathcal{D})\neq \emptyset\right]\overset{N\uparrow\infty}{\longrightarrow} 0\,.
\end{equation}
We call $\alpha_{\mathrm{OGP}}(m)$ the threshold above which $m-$OGP begins to hold. The $m-$OGP has a simple geometric interpretation. It requires that for every $m$-tuple of solutions, there are at least two solutions that are ``close'' or ``far apart''. 
A technical remark is that to rule out stable algorithms, a contradiction argument is usually presented based on a more complicated ``ensemble'' variant of the $m$-OGP, which allows the configurations of the $m$-tuple in the previous definition to be solutions of possibly different instances of the problem. However, as argued in \cite{gamarnik2022algorithms}, the ensemble $m$-OGP and the non-ensemble $m$-OGP are often found to take place at the exact same threshold (we mention here as an example the case of finding the largest independent set in sparse Erdos Renyi random graphs or the symmetric perceptron problem). We refer the reader to~\ref{app:eOGP} for additional details on the non-ensemble version and why we expect that it coincides to OGP in our case. For this reason, from now on we will restrict our analysis to the non-ensemble version. 

Note that if $m'>m$, $m$-OGP implies $m'$-OGP, and therefore $\alpha_{\mathrm{OGP}}(m')\leq \alpha_{\mathrm{OGP}}(m)$. In particular, obstruction to stable algorithms does not require a specific value of $m$, implying that $\alpha_{\mathrm{OGP}}(m)$ represents an upper bound to the values of $\alpha$ that are accessible by stable algorithms, and the best upper bound is the one obtained in the limit $m\rightarrow \infty$ (that is taken after the limit $N\to\infty$). 

Interestingly, in $\text{SBP}_{\kappa}$ it is shown that $m$-OGP, in the limit $m\rightarrow \infty$, appears at $\alpha=O(\kappa^2\log_2{\frac{1}{\kappa}})$ for $\kappa \rightarrow 0$, leading the authors to conjecture that the Bansal and Spencer algorithm is tight up to polylogarithmic factors. In \cite{vafa2025symmetric} their conjecture is proven assuming the worst-case hardness of approximating the shortest vector problem on lattices.

\subsection{The Clonated Free Entropy}
\label{ssec:ClonatedFreeEntropy}
In order to study the $\alpha_{\mathrm{OGP}}(m)$ thresholds, we introduce the following ``clonated'' partition function,
\begin{equation}
	\label{eq::partition_function}
	\mathcal{N}_m (q; \mathcal{D}) \equiv \sum_{\boldsymbol{w}^1,\dots,\boldsymbol{w}^m}\prod_{a = 1}^m\, \mathbb{X}_{\mathcal{D}}(\boldsymbol{w}^a) \prod_{a < b} \delta\left(q-\frac{\boldsymbol{w}^a\cdot\boldsymbol{w}^b}{N}\right).
\end{equation}
This expression counts the number of sets of $m$ solutions (clones), $\boldsymbol{w}_1,\dots,\boldsymbol{w}_m$, conditioned to be at a mutual overlap $q$, given a realization of the disorder $\mathcal{D}$. The indicator,
\begin{equation}
	\label{eq:indicator}
	\mathbb{X}_{\mathcal{D}}(\boldsymbol{w}) = \prod_{\mu=1}^P\Theta\left[ \varphi\left(\frac{1}{\sqrt{N}}\boldsymbol w^*\cdot\boldsymbol x^{\mu} \right) \, \varphi\left(\frac{1}{\sqrt{N}}\boldsymbol w\cdot\boldsymbol x^{\mu} \right)\right]\,,
\end{equation}
where we denote by $\Theta(\bullet)$ the Heaviside step function, enforces that all clones are solutions of a teacher-student problem, with teacher $\boldsymbol{w}^*$. As we shall see, the storage setting can be obtained as a special case of the previous formulas. To shorten the notation, in the following we do not write explicitly the disorder dependence in the partition function.

The partition function \eqref{eq::partition_function} is a random variable that depends on the realization of the disorder. We are interested in studying the typical properties of the clonated system, i.e. the most probable value of \eqref{eq::partition_function}, which in general does not coincide with its average. In the large $N$ limit, the most probable value $\hat{\mathcal{N}}_m(q)$ of $\mathcal{N}_m(q; \mathcal{D})$ is expected to be given by $\hat{\mathcal{N}}_m(q)\sim \exp{(m\,N\hat{\phi}_m(q))}$~\cite{mezard1987spin}, where $\hat{\phi}_m$ is the so-called quenched free entropy \footnote{For a proof of this statement in the ABP see~\cite{talagrand1999}.}:
\begin{equation}
	\label{eq:avfreeEntr}
	\hat{\phi}_m(q)=\lim_{N\rightarrow \infty}\frac{1}{m N}\mathds{E}_{\mathcal{D}}\log{\mathcal{N}_m(q; \mathcal{D})}\,.
\end{equation}
The free-entropy \eqref{eq:avfreeEntr} is upper-bounded by the so-called annealed free-entropy, that is easier to compute analytically:
\begin{equation}
	\hat{\phi}_m(q) \le \phi^{\mathrm{ann}}_m(q) =  \lim_{N \to \infty} \frac{1}{m N} \ln \mathbb{E}_{\mathcal{D}} \mathcal{N}_m (q; \mathcal{D})\,.
\end{equation}
The upper bound follows from the fact that $\mathcal{N}_m(q; \mathcal{D})$ is a non-negative and integer valued random variable, therefore Markov inequality ensures that
\begin{equation}
	\text{Pr}_{\mathcal{D}}[\mathcal{N}_m(q; \mathcal{D})>0] \le  \mathbb{E}_{\mathcal{D}} \mathcal{N}_m(q; \mathcal{D}) = \mathrm{e}^{mN \phi_m^{\mathrm{ann}}(q)}\,.
\end{equation}
Note that if $\phi^{\mathrm{ann}}_m(q) < 0$ for $\alpha$ larger than a certain $\alpha^{\mathrm{\mathrm{ann}}}_m(q)$, then $\text{Pr}[\mathcal{N}_m(q) > 0] \to 0$ for large $N$. This means that for $\alpha \ge \alpha^{\mathrm{ann}}_m(q)$, solutions cannot be found at a fixed mutual overlap $q$, and in particular that the $m$-OGP property holds for $\alpha>\alpha^{\mathrm{ann}}_m(q)$. Based on these inequalities, the $m$-OGP threshold $\alpha_{\mathrm{OGP}}(m)$ can be upper-bounded by $\alpha_{\mathrm{OGP}}^{\mathrm{ann}}(m)=\min_q{\alpha^{\mathrm{ann}}_m(q)}$. The value $\min_q{\alpha^{\mathrm{ann}}_m(q)}$ is computed by looking for the value of the overlap and constraint density such that both the free entropy and its derivative w.r.t.\ $q$ are zero:
\begin{equation}
	\begin{aligned}
		\label{eq:conditionsAlphaOGP}
		\phi_{m}^{\text{ann}}(q)&=0\\
		\frac{\partial}{\partial q}\phi_{m}^{\text{ann}}(q)&=0.
	\end{aligned}
\end{equation}
We observe that the clonated partition function~\eqref{eq::partition_function} and the corresponding entropy~\eqref{eq:avfreeEntr} are conceptually linked to a line of works~\cite{baldassi2015subdominant, baldassi2016unreasonable} that introduce a reweighting of the Gibbs measure of the form $\mathrm{e}^{y \mathcal{S}(\boldsymbol{w}, o)}$, where the term $\mathcal{S}(\boldsymbol{w}, o)$ represents the local entropy of $\boldsymbol{w}$, namely the logarithm of the number of solutions at a fixed overlap 
$o$ from $\boldsymbol{w}$. In our setting, the parameters $m$ and $q$ play roles analogous to $y$ and $o$, respectively. A concrete mathematical connection between the two approaches is discussed in the appendix of~\cite{baldassi2020shaping}. Despite the differences in formulation, both frameworks lead to similar predictions regarding the emergence of an overlap gap property. Furthermore, alternative implicit local entropy biases have been explored in~\cite{baldassi2021unveiling}, providing insights into the onset of hardness transitions in more complex neural network models~\cite{baldassi2022learning}. 

The annealed entropy can be computed using standard techniques (see \ref{app:firstmoment}). Let us introduce the overlaps between the teacher and the clones:
\begin{equation}
	p_a=\frac{1}{N}\sum_{i=1}^Nw_i^{a}w_i^{*}\,. 
\end{equation}
Assuming a symmetric ansatz for the overlaps, $p_a=p$ (see \cref{sec:teacherstudent}), and taking the large $N$ limit, $\phi_m^{\mathrm{ann}}(q)$ becomes:  
\begin{equation}
	\phi_m^{\mathrm{ann}}(q)=\max_{p}\phi_m^{\mathrm{ann}}(q,p)\,,
\end{equation}
where,
\begin{equation}
	\label{eq:annEntrmQP}
	\phi_m^{\mathrm{ann}}(q,p)=G_S^{(m)}(q,p)+\alpha\,G_E^{(m)}(q,p)\,.
\end{equation}
In particular the function $G_S^{(m)}(q,p)$, the ``entropic'' contribution to the annealed free entropy, counts the number of $m$-tuples of vertexes of the hypercube having mutual overlap $q$, and overlap $p$ with the teacher:
\begin{equation}
	\begin{split}
		G_S^{(m)}(q,p)&=\lim_{N\rightarrow \infty}\frac{1}{mN}\log{\sum_{\boldsymbol{w}^1,\dots,\boldsymbol{w}^m}
			\prod_a\delta\left(p-\frac{\boldsymbol{w}^*\cdot\boldsymbol{w}}{N}\right)\,\prod_{a < b} \delta\left(q-\frac{\boldsymbol{w}^a\cdot\boldsymbol{w}^b}{N}\right)}=\\&=\min_{\hat{q},\hat{p}}\left\{-\frac{(m-1)}{2}q\,\hat{q}-p\,\hat{p}-\frac{\hat{q}}{2}+\frac{1}{m}\log{\left(\sum_{k=0}^{m}\binom{m}{k}\mathrm{e}^{(2k-m)\hat{p}+\frac{\hat{q}}{2}(2k-m)^2}\right)}\right\}\,.
	\end{split}
\end{equation}
The function $G_E^{(m)}(q,p)$, the ``energetic'' contribution to the annealed free entropy, counts the probability that an $m$-tuple of vertices with the previous overlap constraints satisfies the perceptron constraints: 
\begin{equation}
	\label{eq:EnergeticTerm}
	G_E^{(m)}(q,p)=\frac{1}{m}\log{\left[2\int Dx \,I\left(\sfrac{p}{\sqrt{q}}\,x,\sqrt{1-\sfrac{p^2}{q}}\right)\left[I\left(\sqrt{q}\,x,\sqrt{1-q}\right)\right]^m\right]}\,.
\end{equation}
The kernel $I(\bullet,\bullet)$ is defined by the activation function $\varphi(\bullet)$ (see \ref{app:firstmoment}). 
Observe that if instead of optimizing over $p$, one fixes $p=0$, the teacher-student partition function reduces to the storage one for every $m$. 

\section{The Storage Problem}
\label{sec:Storage}

In this section, we focus on the storage problem. We start with the computation of the existence of solutions (critical capacity or SAT/UNSAT transition~\cite{Gardner1988,GardnerDerrida1988,KrauthMezard1989,baldassi2019,annesi2024exact}) as a function of $\delta$, see \cref{sec:satthres}. Then in \cref{sec:ogpStor}, we discuss the presence of an OGP for $\alpha\ge \alpha_{\mathrm{OGP}}(\delta)$ as introduced in \cref{ssec:IntroOGP}. In particular, we will see that in the small $\delta$ limit, solutions exist all the way up to $\alpha_c(0)=1$, but OGP holds for any $\alpha>0$. 

\subsection{Satisfiability Threshold}
\label{sec:satthres}

\begin{figure}
	\centering
	\includegraphics[width=0.5\linewidth]{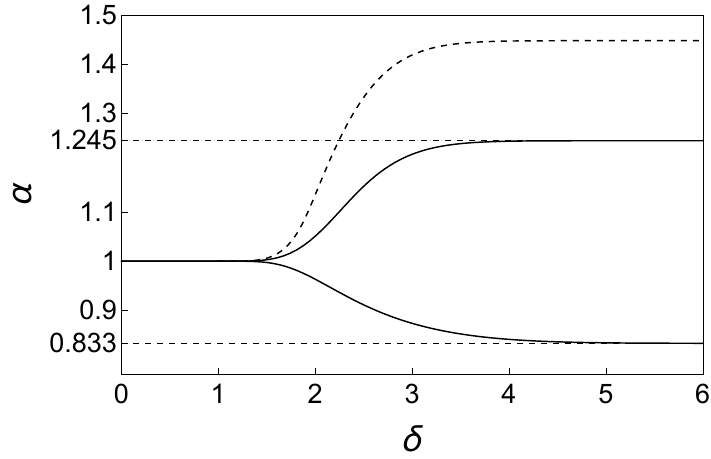}
	\caption{Satisfiability and teacher thresholds of the SWP, respectively $\alpha_c(\delta),\alpha_T(\delta)$, as a function of $\delta$. The continuous line starting from the bottom right represents the RS computation of $\alpha_c(\delta)$. Note that for $\delta\rightarrow \infty$ one recovers the capacity of ABP $\alpha_c^{ABP}\approx 0.833$, for which the RS prediction has been proven to be exact. The continuous line starting from top right represents the teacher threshold $\alpha_T(\delta)$, which in the planted ensemble corresponds to the value of $\alpha$ s.t.\ for $\alpha$ larger the teacher is the unique solution to the problem. For $\delta\rightarrow\infty$ the teacher threshold tends to the ABP one, $\alpha_T^{ABP}\approx 1.245$. The dashed line starting from top right is the annealed computation of the teacher threshold.}
	\label{fig:asatunsatit}
\end{figure}

In order to compute the critical capacity, one has to compute the free entropy of ``one clone'':
\begin{equation}
	\label{eq:FreeEntropyExistence}
	\hat{\phi}_1=\lim_{N\rightarrow\infty}\frac{1}{N}\mathds{E}_{\mathcal{D}}\log{\mathcal{N}_1}\,,
\end{equation}
where, 
\begin{equation}
	\mathcal{N}_1=\sum_{\boldsymbol{w}}\prod_{\mu=1}^P\Theta\left[ y^{\mu} \, \varphi\left(\frac{1}{\sqrt{N}}\boldsymbol w\cdot\boldsymbol x^{\mu} \right)\right]\,,
\end{equation}
is the partition function of one clone, also called \emph{Gardner volume}~\cite{Gardner1988}. The capacity $\alpha_c(\delta)$ is determined using the zero-entropy criterion, which involves finding the smallest value of $\alpha$ such that $\hat{\phi}_1(\alpha)=0$. A simple upper bound of $\hat{\phi}_1$ is provided by the annealed free entropy:
\begin{equation}
	\phi_1^{ann}(\alpha)=\lim_{N\rightarrow\infty}\frac{1}{N}\log{\mathds{E}\mathcal{N}_1}=(1-\alpha)\,\log{2}\,,
\end{equation}
giving an upper bound estimate for the capacity $\alpha_c^{ann}(\delta)=1$ (see \ref{app:annealedStoragecap}). As shown in \ref{app:RSm0existence}, the replica symmetric (RS) ansatz for the free entropy of one clone $\phi^{rs}_1$ can be recovered from the annealed computation with $m$ clones (see Eq.~\eqref{eq:annEntrmQP}) as follows:
\begin{equation}
	\phi^{rs}_1=\max_q\lim_{m\rightarrow 0}\phi^{\mathrm{\mathrm{ann}}}_m(q,0)\,.
\end{equation}
In \cref{fig:asatunsatit} we show the critical capacity estimated with the zero-entropy criterion on the RS free entropy, as a function of $\delta$. We note that this method~\cite{KrauthMezard1989} has been shown to be exact for ABP~\cite{DingSun2019,Huang2024}, which corresponds to taking the SWP in the limit $\delta\rightarrow \infty$ \cite{KrauthMezard1989}. We note that, in the small $\delta$ limit, the RS estimate of the capacity saturates the annealed upper bound. In~\ref{app:SecondMoment}, using a second-moment argument, we show that the first moment estimate of the capacity is correct in this limit.

\subsection{Overlap Gap}
\label{sec:ogpStor}

In \cref{fig:storageOGPa}, we show the annealed estimates of the $\alpha_{\mathrm{\mathrm{OGP}}}(m,\delta)$ thresholds for different values of $m$ and $\delta$. The curves are obtained by the numerical solution of Eqs.~\eqref{eq:conditionsAlphaOGP}. 

\subsubsection{The Small $\delta$ Limit} 
\label{sec:storage_small_delta}

Taking the limit $\delta\rightarrow 0$, it is easy to show that the OGP property holds for any $\alpha>0$. Let us start from the expression of the energetic term of the free entropy
\begin{equation}
	G_E^{(m)}(q)=\frac{1}{m}\log{\left[\int Dx \left[I\left(\sqrt{q}\,x,\sqrt{1-q}\right)\right]^m\right]}.
\end{equation}
The function $I(\bullet,\bullet)$ depends on the specific activation function. In the case of the SWP, we denote it by $I_\delta(\bullet,\bullet)$ (see Eq.~\eqref{eq:KernelSWP}), with the explicit form:
\begin{equation}
	\label{eq:Ideltarew}
	I_\delta(m,\sigma)=\frac{1}{2}-\frac{2}{\pi}\sum_{n=0}^{\infty}\frac{\mathrm{e}^{-\,\pi^2\frac{\sigma^2}{\delta^2}\,\frac{(2n+1)^2}{2}}}{2n+1}\sin{\left((2n+1)\frac{\pi}{\delta}m\right)}.    
\end{equation}
By perturbatively solving Eqs.~\eqref{eq:conditionsAlphaOGP} in the small $\delta$ limit (see~\ref{app:expSmallDelta}) one finds that $\frac{1-q}{\delta^2}\to\infty$, implying that the series in~\eqref{eq:Ideltarew} tends to zero term by term. Therefore, $G_E^{(m)} \to - \log2$ and the free entropy has the following form:
\begin{equation}
	\phi=G_S^{(m)}(q)-\alpha\,\log{2}.
\end{equation}
Since $G_S^{(m)}(q)$ is a monotonic decreasing function of $q$ (see \ref{app:yInfEntropic}), the condition 
\begin{equation}
	0=G_S^{(m)}(1)-\alpha\,\log{2}
\end{equation}
provides the value of $\alpha^{\mathrm{ann}}_{\mathrm{OGP}}(m)$. Since $G_S^{(m)}(1)=\log{2}/m$, we have 
\begin{equation}
	\label{eq::alphaOGP_ann_small_delta}
	\alpha^{\mathrm{ann}}_{\mathrm{OGP}}(m) = \frac{1}{m}   \,, \qquad \delta = 0 
\end{equation}
Therefore, we find that for $\delta\rightarrow 0$, $\alpha_{\mathrm{OGP}}(\infty)=0$, since its annealed upper bound is zero. 

The expansion of Eqs.~\eqref{eq:conditionsAlphaOGP} presented in~\ref{app:expSmallDelta} provides a perturbative estimate $\widetilde{\alpha}^{ann}_{\mathrm{OGP}}(m,\delta)$ of $\alpha^{ann}_{\mathrm{OGP}}(m,\delta)$, which becomes accurate for small $\delta$. In~\ref{app:expSmallDelta} we note that $\widetilde{\alpha}^{ann}_{\mathrm{OGP}}(m,\delta)\geq \alpha^{ann}_{\mathrm{OGP}}(m,\delta)$, and therefore we can use $\min_m \widetilde{\alpha}^{ann}_{\mathrm{OGP}}(m,\delta)$ as an upper bound to the OGP threshold. In~\cref{fig:storageOGPa} we represent the limiting curve $\min_m \widetilde{\alpha}^{ann}_{\mathrm{OGP}}(m,\delta)$ as a function of $\delta$. 

\subsubsection{Failure of the Annealed and RS Ansatz for $\delta \ne 0$}
\label{ssec:failureRS}
\begin{figure}[]
	\centering
	\begin{subfigure}{0.48\textwidth}
		\centering
		\includegraphics[width=\textwidth]{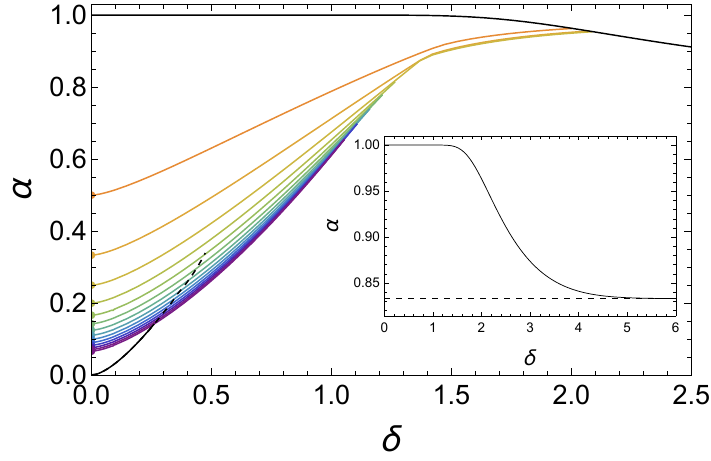}
		\caption{}
		\label{fig:storageOGPa}
	\end{subfigure}
	\hfill
	\begin{subfigure}{0.49\textwidth}
		\centering
		\includegraphics[width=\textwidth]{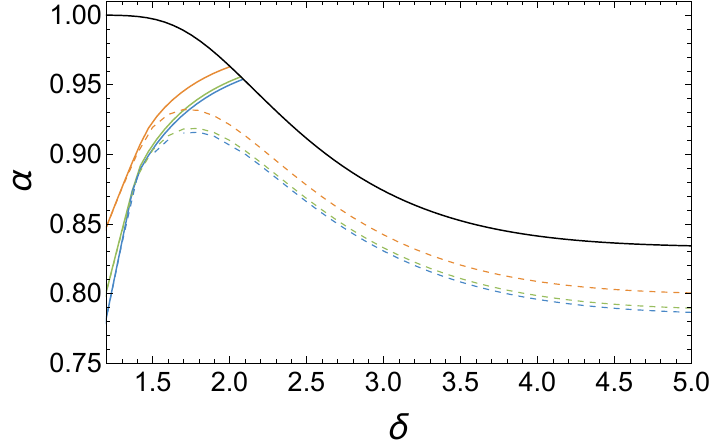}
		\caption{}
		\label{fig:storageOGPb}
	\end{subfigure}
	\caption{(\emph{a}): Storage capacity $\alpha_{sat}$, and annealed $m$-OGP thresholds of the SWP as a function of $\delta$. In particular, for each $\delta$, starting from top to bottom, the colored curves represent the annealed $m$-OGP thresholds for $m=2,\dots,m^{\star}(\delta)$, where $m^{\star}(\delta)=\min{\{15,\text{argmin}_m\{\alpha_{\mathrm{OGP}}(m)\}\}}$. The dots at $\delta=0$ correspond to $1/m$. The black curve at the bottom represents $\min_m \widetilde{\alpha}^{ann}_{OGP}(m,\delta)$ (see Sec.~\ref{sec:ogpStor}). \emph{Inset}: for large $\delta$, $\alpha_{c}$ reaches a plateau (dashed line) corresponding to the storage capacity of the asymmetric binary perceptron $\alpha_{c}^{ABP}\approx 0.833$ (see \cite{KrauthMezard1989}). (\emph{b}): Same as left panel, with the addition of the dashed lines, which represent the replica symmetric corrections to the annealed $m$-OGP estimates for $m=2,3,4$. For small values of $\delta$ the RS predictions collapse on the annealed curves. For large $\delta$, the RS threshold deviate from the annealed ones.}
	\label{fig:storageOGP}
\end{figure}

\begin{figure}[]
	\centering
	\begin{subfigure}{0.49\textwidth}
		\centering
		\includegraphics[width=\textwidth]{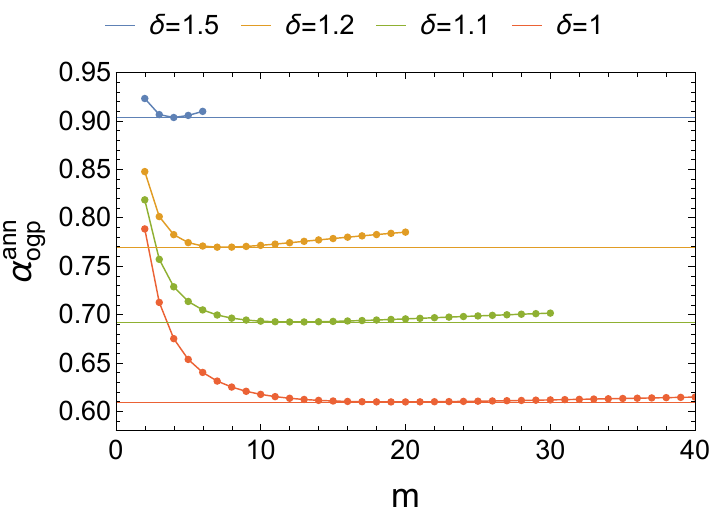}
		\caption{}
		\label{fig:alphaOGP_vs_m_annealed}
	\end{subfigure}
	\hfill
	\begin{subfigure}{0.49\textwidth}
		\centering
		\includegraphics[width=\textwidth]{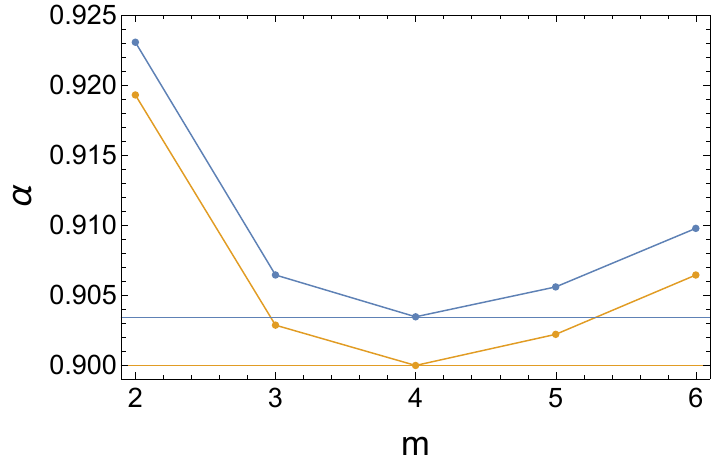}
		\caption{}
		\label{fig:alphaOGP_vs_m_RS}
	\end{subfigure}
	\caption{(\emph{a}): First-moment upper bound of $\alpha_{\mathrm{OGP}}$ in the SWP for different values of $\delta$ in the storage setting. From top to bottom $\delta=1.5,1.2,1.1,1$. The true $\alpha_{\mathrm{OGP}}$ must be a non-increasing function of $m$. The fact that the first moment estimates are non-monotone, implies that the annealed computation cannot be exact. Note that the minimum shifts to larger values of $m$ upon decreasing $\delta$. (\emph{b}): Annealed (upper curve) and replica symmetric (lower dashed curve) estimates of the $\alpha_{OGP}(m)$ thresholds as a function of $m$ for $\delta=1.5$. The replica symmetric ansatz does not cure the unphysical non-monotonicity of the annealed estimates of the thresholds.}
	\label{fig:alphaOGP_vs_m}
\end{figure}

As discussed in \cref{ssec:IntroOGP}, the presence of $m$-OGP implies $m'$-OGP for $m<m'$. Therefore, for $m<m'$, $\alpha_{\mathrm{\mathrm{OGP}}}(m)\leq \alpha_{\mathrm{OGP}}(m')$. However this ordering is not to be respected at the level of the annealed estimates of the thresholds for $\delta \ne 0$, as $\alpha_{\mathrm{OGP}}^{\mathrm{ann}}(m,\delta)$ is a non-monotone function of the number of clones $m$, see \cref{fig:alphaOGP_vs_m_annealed}. This was also observed for $\text{SBP}_{\kappa}$ in~\cite{gamarnik2022algorithms} for finite $\kappa$. Moreover, in \cref{fig:storageOGPa} one can also observe that even taking small values of $m$, the corresponding $\alpha_{\mathrm{ogp}}$ intercepts the SAT/UNSAT transition at a certain value of $\delta$. This is clearly unphysical since in the ABP, that corresponds to $\delta \to \infty$, it has already been observed that an hard algorithmic phase exists. All those inconsistencies are due to the annealed computation of the $m-$OGP thresholds, which only give an upper bound of the true value. We have therefore resorted to a replica symmetric calculation of $\alpha_{\mathrm{OGP}}$, see~\ref{sec:RS_mOGP}.

We show $\alpha^{\mathrm{rs}}_{\mathrm{OGP}}(\delta)$ for $m=2, 3, 4$ in \cref{fig:storageOGPb}. For $\delta \to 0$, the RS estimation coincides with the annealed one, as expected. However, for intermediate and large values of $\delta$ the RS estimation significantly deviates from the annealed one. In particular, for $\delta \to \infty$ and $m=4$ one obtains $\alpha^{\mathrm{rs}}_{\mathrm{OGP}} \simeq 0.7843$ which is close to what was estimated in~\cite{baldassi2015subdominant,baldassi2021unveiling}. 
However, in \cref{fig:alphaOGP_vs_m_RS} we show that for fixed $\delta \ne 0$, even the RS estimate of the OGP threshold suffers the same problem of the annealed approximation, i.e.\ $\alpha_{\mathrm{OGP}}^{\mathrm{rs}}(m)$ is still a non-monotonic function of $m$. This is also signaled by the fact that as soon as $\alpha_{\mathrm{OGP}}^{\mathrm{rs}}(m)$ starts increasing with $m$, the external entropy becomes negative (see \cref{fig:RSComplexity}). 


\subsection{Algorithmic Attack}
\begin{figure}[]
	\centering
	\begin{subfigure}{0.49\textwidth}
		\centering
		\includegraphics[width=\textwidth]{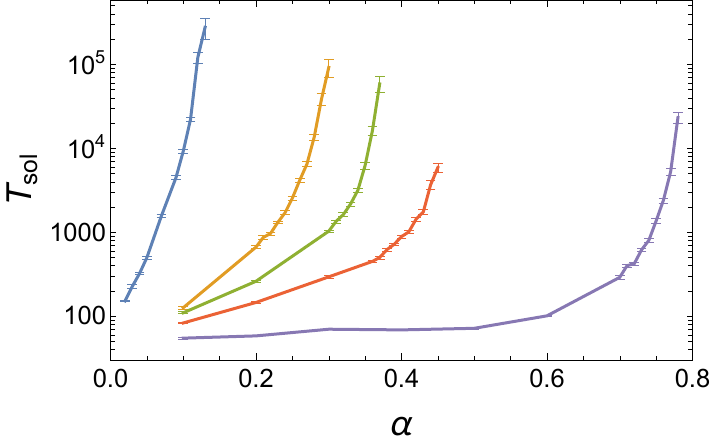}
		\caption{}
		\label{fig:AMPropt}
	\end{subfigure}
	\hfill
	\begin{subfigure}{0.47\textwidth}
		\centering
		\includegraphics[width=\textwidth]{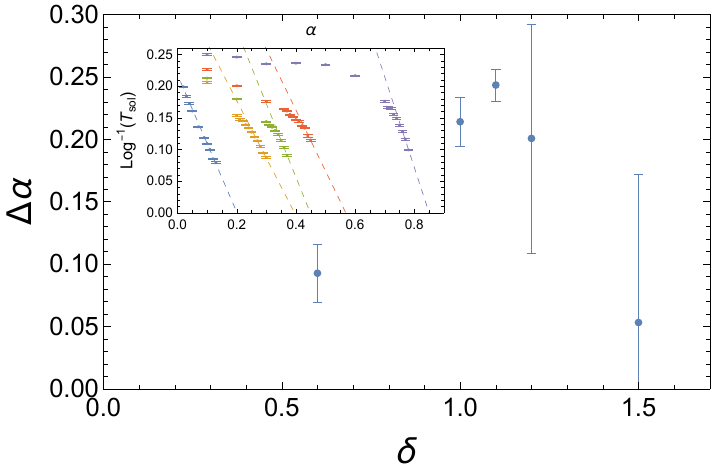}
		\caption{}
		\label{fig:DeltaAlpha}
	\end{subfigure}
	\caption{\emph{(a)}: average number of iterations $T_{sol}$ as a function of $\alpha$, for different values of $\delta$. From right to left $\delta=1.5,1.2,1.1,1,0.6$. The behavior of $T_{sol}$ is compatible with  $T_{sol}\sim \mathrm{e}^{\frac{a}{\alpha_{alg}(\delta)-\alpha}}$, implying that for small $\alpha_{alg}(\delta)-\alpha$, $\log^{-1}{(r_{opt})}\propto \alpha-\alpha_{alg}(\delta)$. This allows to estimate the algorithmic thresholds $\alpha_{alg}(\delta)$ (see the inset of \cref{fig:DeltaAlpha}). Data are obtained averaging over $80$ samples of size $N=4000$. \emph{(b)}: difference $\Delta\alpha(\delta)$ between the optimum value of the RS estimate of the $m$-OGP threshold, namely $\min_{m}\alpha_{OGP}^{rs}(m)$, and the algorithmic thresholds estimated in figure $\emph{(a)}$. \emph{Inset}: $\log^{-1}{(T_{sol})}$ as a function of $\alpha$ for different $\delta$. From right to left $\delta=1.5,1.2,1.1,1,0.6$. Dashed lines are linear fit of the last points.}
	\label{fig:simulazioniampreinforcement}
\end{figure}

In a series of works \cite{braunstein2006learning,Baldassi2015,baldassi2007efficient,baldassi2009generalization,baldassi2015subdominant,baldassi2016unreasonable}, it has been shown that variations of message passing algorithms derived from the cavity method provide powerful solvers for ABP and other CSPs. In this section, we study the behavior of one of these algorithms, called reinforced approximate message passing, ($\text{RAMP}$). In $\text{RAMP}$ each weight $w_i$ of the perceptron is associated with a ``magnetization'' $m_i$, with $-1\leq m_i\leq 1$. Magnetizations are updated according to local rules that depend on an external parameter $r$, called \emph{reinforcement rate}, which has to be adjusted heuristically. We call iteration the set of $O(P N)$ operations needed to update the magnetizations of all sites. After each iteration, the binary vector $\widetilde{\boldsymbol{w}}$, with elements $\widetilde{w}_i=\text{sgn}(m_i)$ is computed and checked to see if it is a solution to the problem. We refer the reader to \ref{app:AMPreinforcementApp} for the details of the algorithm. In \cref{fig:AMPropt} we show that the number of iterations $T_{sol}$ required to find a solution exhibits a critical behavior, namely it diverges in correspondence to an algorithmic threshold $\alpha_{alg}(\delta)$ that depends on $\delta$. In \cref{fig:DeltaAlpha} we show the difference $\Delta\alpha=\min_m\alpha_{OGP}^{rs}(m,\delta)-\alpha_{alg}(\delta)$, between the best replica-symmetric estimate $\min_m\alpha_{OGP}^{rs}(m,\delta)$ of the $m$-OGP threshold and an extrapolation of the algorithmic threshold $\alpha_{alg}(\delta)$, as a function of $\delta$. The shift $\Delta\alpha$ displays a non-monotone behavior. For the largest value of $\delta$ in the dataset, $\delta=1.5$, the shift $\Delta\alpha$ is compatible with zero. This is also the case in the limit of $\delta\rightarrow \infty$, where \cite{baldassi2015subdominant} showed agreement between the estimates of the algorithmic threshold of solvers based on the cavity method and a RS local entropy computation. An intermediate region for $\delta<1.5$ displays positive shifts. However, as we have discussed in \cref{ssec:failureRS}, the RS computation is not exact. Breaking the replica symmetry might be needed to understand this gap. 

\section{The Teacher-Student Problem}
\label{sec:teacherstudent}

In the planted setting, it is natural to distinguish between two algorithmic problems: \emph{optimization}, which asks to find a generic solution, and \emph{learning}, or \emph{recovery}, which asks to find the teacher. Note that in this case, for all values of $\alpha$, there is by definition at least one solution, the teacher $\boldsymbol{w}^*$. Depending on the activation function, one finds a critical value $\alpha_T(\varphi)$, the ``teacher threshold'', s.t.\ for $\alpha\geq \alpha_T(\varphi)$, $\boldsymbol{w}^*$ is the unique solution with high probability, while for $\alpha<\alpha_T(\varphi)$ an exponential number of solution exist. Below $\alpha_T(\varphi)$ the learning task is impossible, since the teacher is indistinguishable from the other solutions. Above $\alpha_T(\varphi)$ optimization and learning are the same problem. 

In \cref{sec:alphateacher} we compute $\alpha_T(\varphi)$ for the SWP. In \cref{sec:ogpTS} we study the presence of an overlap-gap for $\alpha<\alpha_T(\varphi)$. In \cref{sec:recoPlanted}, we discuss the recovery problem. 

\subsection{The Teacher Threshold}
\label{sec:alphateacher}

The teacher threshold can be computed by looking at the one-clone partition function \eqref{eq:FreeEntropyExistence}, with the difference that in the Gardner volume the labels are not random, but produced by the teacher:
\begin{equation}
	\mathcal{N}_1=\sum_{\boldsymbol{w}}\prod_{\mu=1}^P\Theta\left[ \varphi\left(\frac{1}{\sqrt{N}}\boldsymbol w^*\cdot\boldsymbol x^{\mu} \right) \, \varphi\left(\frac{1}{\sqrt{N}}\boldsymbol w\cdot\boldsymbol x^{\mu} \right)\right]\,.
\end{equation}
Using an interpolation method, \cite{barbier2019optimal} proves that if the patterns $\boldsymbol{x}^{\mu}$ have i.i.d.\ elements, and the activation is continuous almost everywhere, the replica-symmetric free entropy for the teacher-student problem is exact. Similarly to the storage case, the replica-symmetric quenched free entropy of the teacher-student problem can be obtained as the limit $m\rightarrow 0$ of the annealed clonated free-entropy, see \ref{app:annealedStoragecap}:
\begin{equation}
	\phi_1^{\mathrm{rs}}=\max_{q,p}\lim_{m\rightarrow 0}\phi_m^{\mathrm{ann}}(q,p)\,.
\end{equation}
Note that in this case, the optimization is over two parameters, the overlaps $p$ and $q$. However, one finds that if the problem is self-averaging, the parameter to fix is just one, since on the fixed point $p=q$. This is a consequence of the so-called Nishimori identities \cite{opper1991calculation,iba1999nishimori,nishimori2001statistical,kabashima2016phase}. One can show that for any $\alpha$, the teacher (i.e.\ $p=1$) is always a maximum with zero entropy. For $\alpha<\alpha_T$, it is only a local maximum, as there exists another fixed point $p<1$ with positive entropy. For $\alpha>\alpha_T$, the point at $p=1$ becomes a global maximum of the free entropy. In \cref{fig:asatunsatit}, we derive $\alpha_T$ as a function of $\delta$ for the SWP. Analogously to the storage case, in the small-$\delta$ region, $\alpha_T$ saturates the lower bound $\alpha_T\geq 1$. An upper-bound is provided by the annealed computation of the free entropy (see \ref{app:annealedTeacherStudent}). 

\subsection{Overlap Gap}
\label{sec:ogpTS}
As we have seen in the previous section, in the teacher-student setting, for $\alpha<\alpha_T(\delta)$,
the replica symmetric computation predicts that an exponential number of configurations of the hypercube satisfies all the constraints of the problem. In this section, we study the presence of OGP in this region. 

\subsubsection{The Small $\delta$ Limit}
In the teacher-student case, to study the annealed bound to the OGP thresholds, one should optimize over all the overlaps between the teacher and the students $p_1,\dots,p_m$ (see \ref{app:firstmoment}). In this section, we discuss this optimization in the limit $\delta\rightarrow 0$. We find, equivalently to the storage case, that the $m$-OGP threshold for the teacher-student problem can be upper-bounded by $1/m$. Let us start from the general expression for the energetic term (see Eq.~\eqref{eq:GenFormGE}):
\begin{equation}
	G_E^{(m)}=\frac{1}{m}\log{\int D_{\Sigma}\,\boldsymbol{u} \prod_{a=1}^m \Theta\left(\varphi_{\delta}(u_0)\,\varphi_{\delta}(u_a)\right)}\,,
\end{equation}
where $D_{\Sigma}\boldsymbol{u}$ is a multivariate Gaussian integration with zero means and covariance matrix $\Sigma$, having the following form:
\begin{equation}
	\label{eq:covmatrix}
	\Sigma=
	\begin{pmatrix}
		1 & p_1 & p_2 & \dots &p_m \\
		p_1 & 1 & q & \dots & q  \\
		p_2 & q & 1 & \dots & q \\
		\vdots & \vdots & \vdots & \ddots & \vdots \\
		p_m & q & q & \dots & 1 \\
	\end{pmatrix}\,.
\end{equation}
If $\Sigma$ is full-rank, the Gaussian integration is smooth and in the limit $\delta\rightarrow 0$ one simply has:
\begin{equation}
	\label{eq::Ge_small_delta}
	G_E^{(m)}\rightarrow -\log{2}\,.
\end{equation}
This follows from the fact that the theta functions tessellate the $(m{+}1)$-dimensional integration space into boxes of side $\delta$, alternating between values $1$ and $0$. Since the Gaussian measure is regular, in the limit $\delta \to 0$ the integral reduces to the average value of the boxes, which equals $1/2$. 

Let us take $q<1$. Depending on the value of the overlaps $p_1,\dots,p_m$ there can be \emph{at most} one zero mode, since the vectors $(1,q,\dots,q),(q,1,\dots,q),\dots,(q,q,\dots,1)$ for $q<1$ are linearly independent. Suppose that there is one zero mode. In this case the Gaussian integration is constrained by a condition of the following form:
\begin{equation}
	\label{eq:lincomb}
	u_0=c_1\,u_1+c_2\,u_2+\dots+c_m\,u_m\,,
\end{equation}
for some coefficients $c_1,\dots,c_m$. Note that there cannot be a zero coefficient in front of $u_0$ since the rows corresponding to the students are linearly independent. Note also that the $u$'s have unitary variance:
\begin{equation}
	\langle u_0^2\rangle=\langle u_1^2\rangle=\dots=\langle u_m^2\rangle=1\,,
\end{equation}
where we denoted with angle brackets the expectation according to a multivariate Gaussian with zero mean and covariance matrix $\Sigma$. Squaring Eq.~\eqref{eq:lincomb}, and computing the expectation, we find the following relation:
\begin{equation}
	\label{eq:sqa2}
	1=\sum_ac_a^2+2\,q\sum_{a<b}c_a\,c_b\,.
\end{equation}
The request that $\Sigma$ is a covariance matrix implies a further constraint. Indeed, it must be possible to write the elements of $\Sigma$ as $\Sigma_{ab}=\frac{1}{N}\sum_i\tau_a^i\,\tau_b^i$, for some $\tau_a^i=\pm 1$. This constraint is embodied in the entropic contribution to the free entropy, which can be written as (see \cref{eq:APP_Gs})

\begin{equation}
	G_S^{(m)}(q,p_1,\dots,p_m)= \frac{1}{mN}\log{\frac{1}{2^N}\sum_{\boldsymbol{w}_0\cdots \boldsymbol{w}_m}\prod_{a<b}\delta\left(q-\frac{\boldsymbol{w}_a\cdot\boldsymbol{w}_b}{N}\right)\,\prod_a\delta\left(p_a-\frac{\boldsymbol{w}_0\cdot\boldsymbol{w}_a}{N}\right)}
\end{equation}
Any $\Sigma$ not fulfilling the above constraint would lead to an infinitely negative $G_S$.  
As we will be interested in the $N\to\infty$ limit, the quantized nature of the value of $q, p_1, ...$ is lost, but other implications of the constraint remain strong even in this limit\footnote[3]{We note here that the requirement that $\Sigma$ can be written as $\Sigma_{ab}=\frac{1}{N}\sum_i\tau_a^i\,\tau_b^i$ is stronger than positive definitiveness. For example it is easy to check that the matrix with $m=2$
	\begin{equation*}
		\begin{pmatrix}
			1 & p_1 & p_2 \\
			p_1 & 1 & q \\
			p_2 & q & 1
		\end{pmatrix}
	\end{equation*}
	is not acceptable whenever $p_2 \le q + p_1 - 1$ (as it violates triangular inequality), but can be positive definite.}. Namely, if $\tilde{\boldsymbol{u}}$ is an eigenvector of $\Sigma$ associated with a zero eigenvalue, one has:
\begin{equation}
	\label{eq:conditionCorrMatrix}
	0=\tilde{\boldsymbol{u}}^T\Sigma\, \tilde{\boldsymbol{u}}=\sum_{ab}\tilde{u}_a\tilde{u}_b\left(\frac{1}{N}\sum_{i}\tau_a^i\tau_b^i\right)=\frac{1}{N}\sum_i\left(\sum_a\tau_a^i\,\tilde{u}_a\right)^2\,.
\end{equation}
Hence, there must exist at least one binary sequence 
$\tau_1,\dots,\tau_m\in\{\pm1\}^m$ s.t.\ 
\begin{equation}
	\label{eq:conditionCorrMatrix_2}
	\sum_a\tau_a^i\,\tilde{u}_a=0.
\end{equation}
From Eq.~\eqref{eq:lincomb}, the eigenvector associated with the zero mode can be written up to a normalization as $\tilde{u}=(1,-c_1,-c_2,\dots,-c_m)$. Therefore Eq.~\eqref{eq:conditionCorrMatrix_2} implies that there must exist at least one binary sequence 
$\tau_1,\dots,\tau_m\in\{\pm1\}^m$ s.t.\ 
\begin{equation}
	\label{eq:condCovar}
	1=\tau_1\,c_1+\tau_2\,c_2+\dots+\tau_m\,c_m\,.
\end{equation}
If there is just \emph{one} non-zero $c_a$, say $c_1$, then the previous condition implies $c_1=\pm 1$. Using also the constraint \eqref{eq:lincomb} one finds:
\begin{equation}
	\label{eq:lru0ua}
	\langle u_0\, u_a\rangle=\pm \langle u_1\,u_a\rangle\,,
\end{equation}
Equation \eqref{eq:lru0ua} can be satisfied if and only if the teacher coincides with one of the students (the first one in this example): $p_1=1$, $p_2=p_3=\dots=p_m=q$. In this case, the energetic term becomes equal to the one of the storage case with $m$ clones. The entropic term $G_S$ is equivalent to a storage problem where one of the clones is constrained to a random vertex of the hypercube. Therefore, the overall free entropy must be at most equal to the one of the storage case with $m$ clones. But as we have seen in the previous section, in the storage case, the annealed estimate of the $m$-OGP is upper-bounded by $1/m$, so the same bound holds also in the present case.

Suppose now that there are at least two non-zero coefficients. Squaring \cref{eq:condCovar}, and subtracting \cref{eq:sqa2} we get the following condition:
\begin{equation}
	\sum_{a<b}(\tau_a\tau_b-q)\,c_ac_b=0\,,
\end{equation}
that is equivalent to ask that the matrix
\begin{equation}
	S=\boldsymbol{\tau}\boldsymbol{\tau}^T-(\mathds{1}+q\,J)
\end{equation}
is not invertible, where we denoted by $J$ the matrix that has zeros on the diagonal and ones elsewhere. Note that $S$ is a rank-one perturbation of a full-rank matrix, therefore using the Sherman-Morrison formula, the non invertibility condition can be rewritten as:
\begin{equation}
	\label{eq:conditionEigenvalue1}
	1-\boldsymbol{\tau}^T(\mathds{1}+q\,J)^{-1}\boldsymbol{\tau}=0\,.
\end{equation}
The inverse $(\mathds{1}+q\,J)^{-1}$ can be computed analytically:
\begin{equation}
	(\mathds{1}+q\,J)^{-1}=\frac{1}{(1-q)\left(1+(m-1)\,q\right)}\bigg(\big(1+(m-2)\,q\big)\,\mathds{1}-q\,J\bigg)\,,
\end{equation}
allowing to rewrite condition \eqref{eq:conditionEigenvalue1} as follows:
\begin{equation}
	\label{eq:conditionEigenvalue}
	\sum_a\tau_a=\frac{1}{\sqrt{q}}\sqrt{(1+(m-1)\,q)(m-1+q)}\,.
\end{equation}
However this equation is not solvable by any $\tau_1,\dots,\tau_m\in \{\pm1\}^m$ for any $m$ unless $q=1$. Indeed the right-hand side of Eq.~\eqref{eq:conditionEigenvalue} is always greater than or equal to $m$ (with equality attained for $q=1$), whereas the left-hand side can only be an integer in the range $-m, -m+2, \dots, m-2, m$. This means that the only correlation matrices satisfying Eq.~\eqref{eq:conditionEigenvalue1} are those produced by configurations that differ from the all ones vector by at most $o(m)$ bits. Therefore if $q<1$ it is not possible to choose more than one non-zero $c_a$, and in the limit $\delta \to 0$ the only possibility is that the energetic term reduces to~\eqref{eq::Ge_small_delta}. By the same argument exposed in the storage case in \cref{sec:storage_small_delta}, this implies that for $\delta \to 0$ the $m$-OGP threshold is given by $1/m$ as in Eq.~\eqref{eq::alphaOGP_ann_small_delta} and vanishes in the large $m$ limit. 

\begin{figure}[]
	\centering
	\begin{subfigure}{0.49\textwidth}
		\centering
		\includegraphics[width=\textwidth]{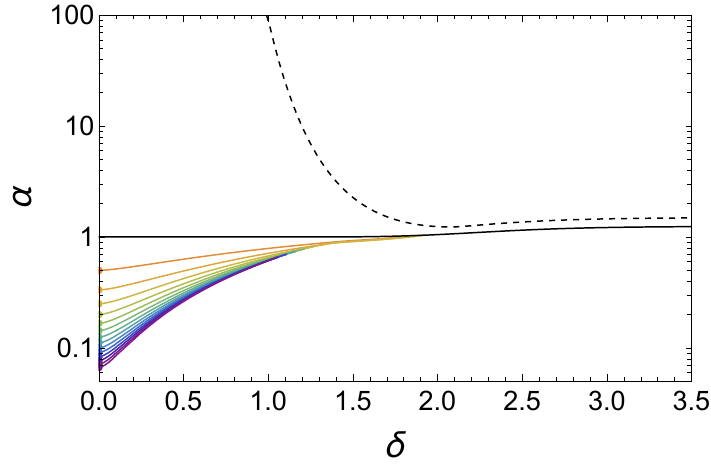}
		\caption{}
		\label{fig:teacherstudentOGPa}
	\end{subfigure}
	\hfill
	\begin{subfigure}{0.49\textwidth}
		\centering
		\includegraphics[width=\textwidth]{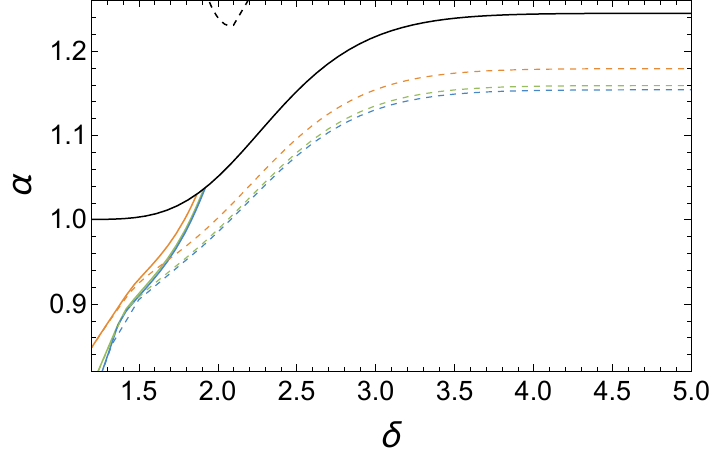}
		\caption{}
		\label{fig:teacherstudentOGPb}
	\end{subfigure}
	\caption{(\emph{a}):
		Starting from the top right, the dashed line represents the hard-easy transition (see Sec.~\ref{sec:recoPlanted}), above which AMP is able to fully reconstruct the teacher, as a function of $\delta$. For $\delta\rightarrow \infty$ the hard-easy tends to $\alpha_{r}\approx 1.49$, which is the spinodal transition of the planted ABP. Moreover notice how this line is non-monotonic in $\delta$, for an intuitive explanation see~\ref{app::non-monotonicity_hard-easy}. The black continuous line represents the teacher threshold $\alpha_{T}$, above which the teacher is the unique solution, as a function of $\delta$. The colored lines are the annealed upper bounds of the $m$-OGP, as a function of $\delta$. In particular, going from top to bottom, $m=2,\dots,m^{\star}(\delta)$, where $m^{\star}(\delta)=\min{\{15,\text{argmin}_m\{\alpha_{\mathrm{OGP}}(m)\}\}}$. The dots at $\delta=0$ correspond to $1/m$. (\emph{b}): The dashed line represents the replica symmetric corrections to the annealed $m$-OGP estimates for $m=2,3,4$. From small values of $\delta$ the RS predictions collapse on the annealed curves. For large $\delta$, the RS threshold deviate from the annealed ones.}
	\label{fig:teacherstudentOGP}
\end{figure}

\subsubsection{The $\delta \ne 0$ Case}

When $\delta$ is different from zero, the OGP transition lines retain aspects similar to those examined in the storage setting. In particular, for $\delta$ sufficiently large, the annealed approximation gives estimates of $\alpha_{\mathrm{OGP}}$ that are above $\alpha_T$, see \cref{fig:teacherstudentOGPa}. This is clearly wrong, as it has already been shown in the ABP~\cite{baldassi2015subdominant} that one can find a generic solution only strictly below $\alpha_T$. Using an RS ansatz solves this problem (see \cref{fig:teacherstudentOGPb}), and in particular for $\delta \to \infty$ and $m=4$ one obtains $\alpha^{\mathrm{rs}}_{\mathrm{OGP}} \simeq 1.15$, again similar to the value reported in~\cite{baldassi2015subdominant}. However, as in the storage case, the RS ansatz is not enough to correctly compute the value of $\alpha_{\mathrm{OGP}}$ for large $m$.

\subsection{Recovery of the Planted Signal}
\label{sec:recoPlanted}
In this section we address the recovery problem, which consists in the inference of the planted solution. It can be useful to distinguish between \emph{weak recovery}, which corresponds to inferring a finite fraction of bits of the teacher, from full recovery, which corresponds to inferring all the bits. Full recovery below $\alpha_T(\varphi)$ is \emph{impossible}, since there is an exponential number of solutions and the teacher is indistinguishable from them. Above $\alpha_T(\varphi)$, full recovery is instead information-theoretically possible. Indeed, since the planted solution is the unique solution, an exhaustive search would find it. Interestingly, depending on the activation function and the domain of definition of the weights, it might happen that the best efficient algorithms require a fraction of patterns \emph{strictly} larger than $\alpha_T(\varphi)$ in order to fully recover the teacher; namely they display a statistical-to-computational gap. This is the case, for example, both in planted $\text{SBP}_{\kappa}$ and in ABP \cite{baldassi2015subdominant,barbier2019optimal}. From the static point of view, in hard inference problems, there is a region above $\alpha_T(\varphi)$ characterized by a coexistence between a thermodynamically stable phase, described by the fixed point of the replica-symmetric free entropy associated with the planted solution, and a so-called \emph{metastable} phase, associated with another locally stable fixed point with a lower free entropy. The spinodal transition $\alpha_r$ is defined as the value of $\alpha$ s.t.\ for $\alpha>\alpha_r$ the metastable fixed point becomes locally unstable. In general, no known efficient algorithm can recover the planted signal below $\alpha_r$. An exception is represented by the peculiar case of planted $k$-XOR-SAT, where the metastable phase extends all the way up to $\alpha=O(N^{k/2})$ \cite{feldman2015complexity}, but due to its linear structure, the problem can be solved in polynomial time by Gaussian elimination. In ``hard'' inference problems, the spinodal transition is believed to represent a fundamental barrier to efficient algorithms. In particular, belief propagation (BP) and variations of spectral methods saturate the spinodal threshold \cite{zdeborova2016statistical}. Numerical evidences support the same for a replicated version of Monte Carlo, called replicated simulated annealing \cite{angelini2023limits}. In binary perceptron problems, above the spinodal transition, BP and Approximate Message Passing (AMP) (which for these problems is equivalent to BP in the large $N$ limit) converges to the planted solution. Given a realization of the dataset, in AMP one writes down iterative equations for single-variable marginals of the uniform measure over all solutions of the perceptron (see \ref{app:AMPreinforcementApp}). Since we are considering the case of binary weights $w_i$, each marginal $\mu_i(w_i)$ can be parameterized by a single number, e.g.\ the magnetization $m_i=\sum_i\mu_i(w_i)\,w_i$. The $m_i$ depends on the site $i$. It turns out that this dependence can be characterized statistically in the large $N$ limit. Indeed, it is possible to write closed equations (state evolution) for the probability distributions of the magnetizations iterated by AMP. A remarkable fact is that the fixed points of the state evolution equations are in correspondence with the fixed points of the replica symmetric free entropy. For this reason, the properties of the metastable fixed point that one finds from the analysis of the RS free entropy is crucial for the behavior of AMP. Indeed, as we are going to see, an initialization of the AMP messages at zero overlap with the planted signal (non-informed initialization), leads in the hard phase to the metastable fixed point. Note that AMP returns marginals, not configurations of the hypercube. However, in perceptrons the fixed point associated with the planted signal has overlap $p=1$ with the teacher. This is a consequence of the fact that the teacher is an isolated solution. In other words, on this fixed point the magnetizations are all polarized to $\pm 1$, namely they correspond to a configuration of the hypercube (the teacher), and therefore AMP is an actual solver for the recovery problem. 

Let us discuss how to compute the spinodal transition. We call $\phi(p,\alpha)$ the replica symmetric free entropy on the Nishimori line, where $p=q$ (see \ref{app:RSm0existence}). In order to identify $\alpha_r$, one should find $p^*<1$ such that:
\begin{equation}
	\label{eq:HardEasy}
	\begin{cases}
		\phi'(p^*;\alpha_r)=0\,,\\
		\phi''(p^*;\alpha_r)=0\,.
	\end{cases}
\end{equation}
The first equation asks $p^*$ to be a fixed point. The second is the marginality requirement of the fixed point that identifies $\alpha_r$. Equations \eqref{eq:HardEasy} can be solved numerically. In the case of the SWP, they lead to the predictions shown in \cref{fig:teacherstudentOGPa}.

A particularly interesting case is when the metastable fixed point is at zero overlap with the teacher, namely when $p^*=0$. In this case, we say that the metastable fixed point is \emph{non-informative}, because it does not carry information about the teacher. As long as the non-informative fixed point is stable, AMP is equivalent to random guessing the bits of the teacher. A necessary condition for doing better is that the non-informative fixed point destabilizes. In this case, there are two possibilities. If there is another (metastable) fixed point on the Nishimori line that is different from the one associated with the teacher, then weak recovery becomes possible. If the teacher is the only other fixed point, then the teacher can be fully recovered. Let us expand the free entropy around $p=0$ along the Nishimori line. For the entropic term one gets:
\begin{equation}
	G_S(p)\approx \log{2}-\frac{p^2}{4}\,.
\end{equation}
For the energetic term:
\begin{equation}
	\label{eq:ExpansionGEnonInfo}
	G_E(p)\approx g_0+g_1\,p+g_2\,\frac{p^2}{4}\,.
\end{equation}
The coefficients $g_0,g_2,g_2$ read:
\begin{equation}
	g_0=\sum_\tau H_0(\tau)\log{H_0(\tau)}\,,\quad g_1=\frac{1}{2}\sum_{\tau}\frac{H_1^2(\tau)}{H_0(\tau)}\,,\quad g_2=\sum_{\tau}\frac{\big(H_1(\tau)-H_0(\tau)H_2(\tau)\big)^2}{H_0^3(\tau)}\,,
\end{equation}
where 
\begin{equation}
	\label{eq:definitionH0H1H2}
	H_0(\tau)=\int Dh\,\Theta(\tau\varphi(h))\,,\quad H_1(\tau)=\int Dh\,h\,\Theta(\tau\varphi(h))\,,\quad H_2(\tau)=\int Dh\,(h^2-1)\,\Theta(\tau\varphi(h))\,.
\end{equation}
are the first Hermite polynomials. See, e.g., Refs.~\cite{damianComputationalStatisticalGapsGaussian2024,maillardPhaseRetrievalHigh2020,mondelliFundamentalLimitsWeak2019,troianiFundamentalComputationalLimits2024} for representative works analyzing the link between weak recovery by efficient algorithms and the Hermite decomposition of the activation. 

If the activation function is balanced, that is, if
\begin{equation}
	\int Dh\,\varphi(h)=0,
\end{equation}
then $H_0(\tau)=1/2$, and therefore $g_0=-\log{2}$. The previous formulas tell us that in order for the metastable fixed point to be non-informative one should have $H_1(\tau)=0$ (because one should have $g_1=0$). If the activation is even, this condition is automatically satisfied. If $g_1=0$, then the non-informative fixed point is stable up to $\alpha_r=1/g_2$. If $g_2=0$, then the non-informative point is always stable for $\alpha=O(1)$, meaning that AMP cannot do better than random guessing for $\alpha=O(1)$. For example, in the case of the SWP, for $\delta\rightarrow 0$, we have $g_1,g_2\rightarrow 0$.  

In what follows, we give a physical interpretation of the Hermite decomposition and show that a simple modification of the SWP reproduces the divergence of the AMP recovery threshold without requiring to take an infinite-frequency limit.

\subsubsection{The Spinodal Transition in the Reversed Wedge Perceptron}
\begin{figure}[]
	\centering
	\begin{subfigure}{0.49\textwidth}
		\centering
		\includegraphics[width=\textwidth]{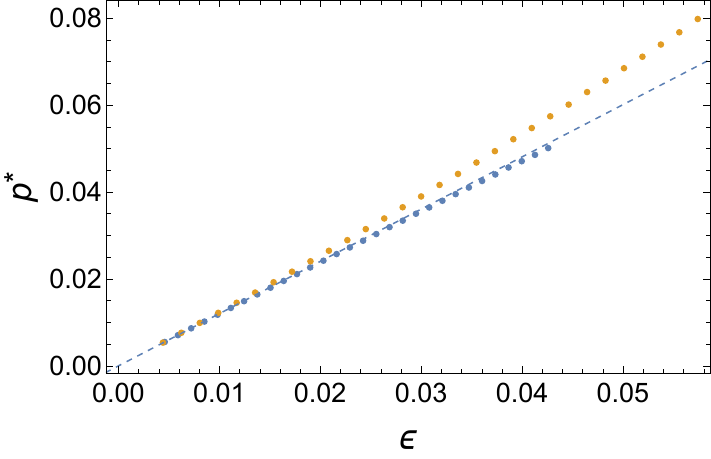}
		\caption{}
		\label{fig:fig1}
	\end{subfigure}
	\hfill
	\begin{subfigure}{0.49\textwidth}
		\centering
		\includegraphics[width=\textwidth]{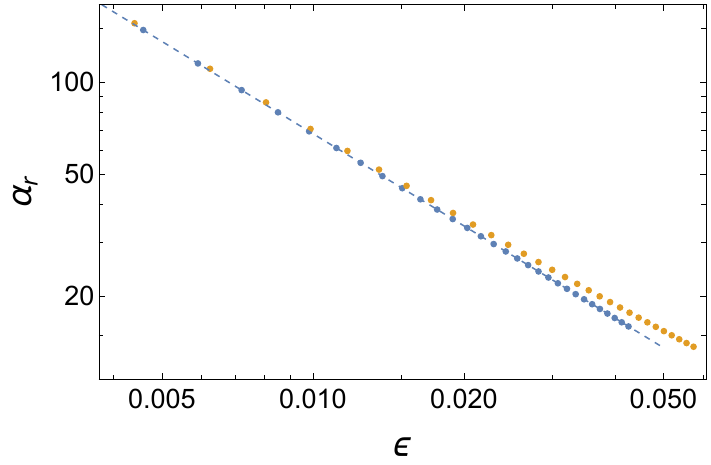}
		\caption{}
		\label{fig:fig2}
	\end{subfigure}
	\caption{(\emph{a}): scaling of the overlap $p^*$ with the teacher computed on the metastable fixed point, as a function of the distance $\varepsilon=|\gamma_{\star}-\gamma|$ from $\gamma_{\star}$. Points are computed by solving numerically the RS saddle-point equations. The upper (orange) points correspond to $\gamma<\gamma_{\star}$, while the lower (blue) points correspond to $\gamma>\gamma_{\star}$. The dashed line is the leading order term $p^*\approx (\log{2})^{-1/2}\varepsilon$. (\emph{b}): scaling of the Hard-Easy transition close to $\gamma_{\star}$ as a function of $\varepsilon=|\gamma_{\star}-\gamma|$. Points are computed by solving numerically the RS saddle-point equations.  The upper (orange) points correspond to $\gamma<\gamma_{\star}$, while the lower (blue) points correspond to $\gamma>\gamma_{\star}$. The dashed line is the leading order term $\alpha_r(\gamma)\approx \pi/8\,(\log{2})^{-3/2}\,\varepsilon^{-1}$.}
	\label{fig:comparison_epsilon_small_rwp}
\end{figure}

\begin{figure}[]
	\centering
	\begin{subfigure}{0.53\textwidth}
		\centering
		\includegraphics[width=\textwidth]{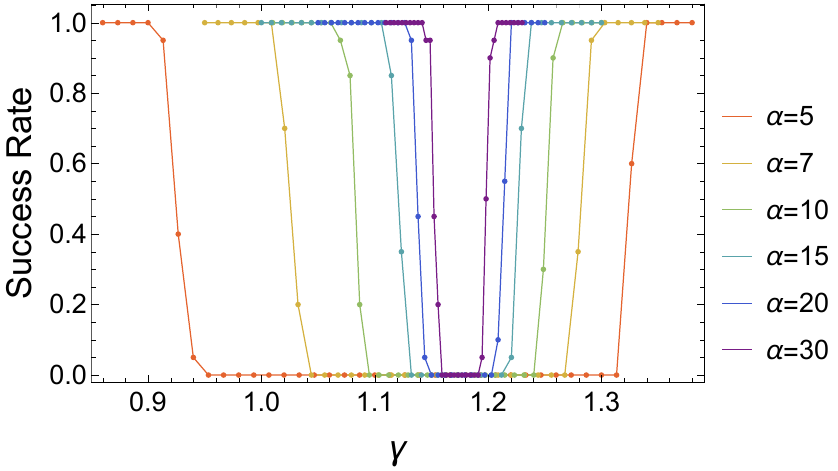}
		\caption{}
		\label{fig:amp_success_rate}
	\end{subfigure}
	\hfill
	\begin{subfigure}{0.45\textwidth}
		\centering
		\includegraphics[width=\textwidth]{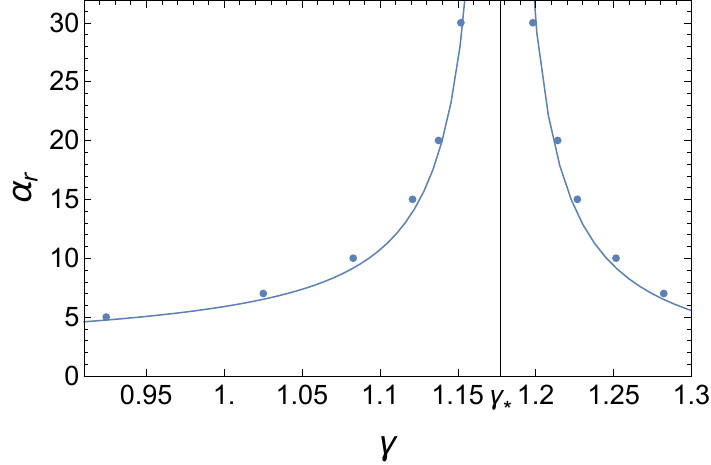}
		\caption{}
		\label{fig:amp_spinodal_transition}
	\end{subfigure}
	\caption{(\emph{a}): Success rate of AMP in finding the planted signal when using the RWP activation function in~\eqref{eq:RWPdef}. We plot the success rate as a function of $\gamma$, for different values of $\alpha$. Each point is averaged over $20$ independent samples of size $N=5\times 10^3$. (\emph{b}): Hard-Easy transition $\alpha_r$ as a function of $\gamma$. The transition diverges at $\gamma=\gamma_{\star}=\sqrt{2\log{2}}$. The continuous line is computed by solving numerically the RS saddle-point equations. The points are numerical simulations of AMP obtained by averaging over $20$ independent samples of size $N=5\times 10^3$.}
	\label{fig:rwp_success_rate}
\end{figure}

In order to set $g_1=g_2=0$, it is sufficient to consider a truncated version of the square wave, defined as follows:
\begin{equation}
	\label{eq:RWPdef}
	\varphi(h)=\text{sgn}\big((h-\gamma)\,h\,(h+\gamma)\big)\,.
\end{equation}
The activation~\eqref{eq:RWPdef} is known as the reversed wedge perceptron (RWP)~\cite{watkin1992,boffetta1993,engel1995,engel2001statistical}. Since~\eqref{eq:RWPdef} is odd, we automatically have $H_2(\tau)=0$. Putting $H_1=0$, one finds $\gamma_{\star}=\sqrt{2\log{2}}$. Let us call $\varepsilon=|\gamma-\gamma_{\star}|$. The following scalings are found for small $\varepsilon$:
\begin{equation}
	\label{eq:scalings}
	p^{*}\approx U\,\varepsilon,\quad \alpha_r\approx A\, \varepsilon^{-1}\,,
\end{equation}
therefore, in the limit $\gamma\rightarrow\gamma_{\star}$ the metastable state does not destabilize for $\alpha=O(1)$ within a RS ansatz. Using the scalings~\eqref{eq:scalings} in equations~\eqref{eq:HardEasy}, where the expression for the free entropy is given in \ref{app:RSm0existence}, and expanding for small $\varepsilon$, it is possible to compute $U,A$ defined in Eq.~\eqref{eq:scalings}. In order to fix $U$ and $A$, one has to expand the energetic term up to $p^3$. One finds:
\begin{equation}
	\label{eq:GEexpanded}
	G_E(p)=a_0(\varepsilon)+a_1(\varepsilon)\,p+a_2(\varepsilon)\,p^2+a_3(\varepsilon)\,p^3\,,
\end{equation}
where,
\begin{equation}
	a_0(\varepsilon)\approx -\log{2},\quad a_1(\varepsilon)\approx \frac{2\log{2}}{\pi}\varepsilon^2\,,\quad a_2(\varepsilon)\approx \frac{4\,(\log{2})^2}{\pi^2}\varepsilon^4\,,\quad a_3(\varepsilon)\approx \frac{2\,(\log{2})^2}{3\pi}\,.
\end{equation}
Therefore equations~\eqref{eq:HardEasy} become,
\begin{equation}
	\begin{cases}
		-p+2\alpha\left(a_1(\varepsilon)+2\,a_{2}(\varepsilon)\,p+3\,a_3(\varepsilon)\,p^2\right)=0\,,\\
		-1+2\alpha\left(2\,a_2(\varepsilon)+6\,a_3(\varepsilon)\,p\right)=0\,.
	\end{cases}
\end{equation}
At this point using ansatz~\eqref{eq:scalings}, we get a closed equation for $U,A$:
\begin{equation}
	\begin{cases}
		-U+2A\left(a_1(\varepsilon)+3\,a_3(\varepsilon)\,U^2\right)=0\,,\\
		-1+12\,A\,a_3(\varepsilon)\,U=0\,.
	\end{cases}
\end{equation}
Note that the term proportional to $a_2$ does not contribute since it is $O(\varepsilon^4)$. Therefore, one finds:
\begin{equation}
	\label{eq:scalingspstarspinodalrwp}
	p^{*}\approx (\log{2})^{-1/2}\,\varepsilon,\quad \alpha_{r}\approx \frac{\pi}{8}(\log{2})^{-3/2}\varepsilon^{-1}\,.
\end{equation}
In \cref{fig:comparison_epsilon_small_rwp} we show a comparison between the scalings of Eq.~\eqref{eq:scalingspstarspinodalrwp} and the numerical solution of Eq.~\eqref{eq:HardEasy} for small $\varepsilon$. In \cref{fig:amp_success_rate,fig:amp_spinodal_transition} we compare numerical experiments performed with AMP with the analytical prediction for the spinodal transition for different values of $\gamma$.

\subsubsection{Bias Detection in the Planted Ensemble}
\label{sec:ComputationDistinguishers}

Unlike the storage problem, in the teacher-student setting, the disorder exhibits correlations. Indeed, due to the presence of the teacher, patterns sharing the same label have a non-trivial geometric structure. Interestingly, these correlations determine the coefficients of the expansion \eqref{eq:ExpansionGEnonInfo} of the free entropy around the non-informative point, allowing us to interpret the divergence of the spinodal transition observed in the last section. To detect these biases, it is sufficient to compute simple observables conditioned on one label. An example is the barycenter:
\begin{equation}
	\boldsymbol{x}^{(\pm)}=\frac{1}{P}\sum_{\mu}\Theta(\pm\,y^{\mu})\,\boldsymbol{x}^{\mu}\,,
\end{equation}
where $\boldsymbol{x}^{(+)}$ ($\boldsymbol{x}^{(-)}$) is the barycenter of the patterns associated with a positive (negative) label. In the storage problem, one simply has that $\mathds{E}x^{(+)}_i=\mathds{E}x^{(-)}_i=0$ for all $i$'s, since the $y^{\mu}$'s are i.i.d.\ Rademacher. In the teacher-student we have to compute:
\begin{equation}
	x^{(y)}_k=2^P\,\mathds{E}_{\{\boldsymbol{x^{\mu}}\}}\left[\prod_{\mu}\Theta\left(y\,\varphi\left(\frac{1}{\sqrt{N}}\sum_ix^{\mu}_iw_i^{*}\right)\right)\frac{1}{P}\sum_\mu x^{\mu}_k\right]\,,
\end{equation}
where the normalization $2^P$ comes from the fact that each pattern has probability $1/2$ of being associated with label $y$. Carrying the expected value inside the sum we get,
\begin{equation}
	x^{(y)}_k=\frac{2}{P}\sum_{\mu}\,\mathds{E}_{\{\boldsymbol{x^{\mu}}\}}\left[\Theta\left(y\,\varphi\left(\frac{1}{\sqrt{N}}\sum_ix^{\mu}_iw_i^{*}\right)\right)\,x^{\mu}_k\right]=2\int D_{\Sigma}(u,v)\,\Theta\left(y\,\varphi(u)\right)\,v\,,
\end{equation}
where,
\begin{equation}
	\Sigma=
	\begin{pmatrix}
		1 & w_k^*/\sqrt{N}  \\
		w_k^*/\sqrt{N} & 1 
	\end{pmatrix}\,.
\end{equation}
Therefore in the end we get,
\begin{equation}
	x^{(y)}_k\approx 2\,H_1(y)\,\frac{w_k^*}{\sqrt{N}}\,,
\end{equation}
where $H_1$ is the same function defined in Eq.~\eqref{eq:definitionH0H1H2} determining the coefficient $g_1$ of the expansion \eqref{eq:ExpansionGEnonInfo}. In the case of the RWP of the previous section, $H_1(y,\gamma)$ depends also on the parameter $\gamma$ and is given by:
\begin{equation}
	H_1(y,\gamma)=y\frac{2\,\mathrm{e}^{-\gamma^2/2}-1}{\sqrt{2\pi}}\,.
\end{equation}
The previous formulas tell us that the barycenter is generally biased towards the teacher. If one chooses the activation s.t.\ $H_1=0$ the barycenter becomes non-informative, and also the first term $g_1$ of the expansion of the free entropy vanishes, implying that the point at zero overlap with the teacher in the state evolution equations becomes a fixed point. 

Let us study another observable that is related to the second coefficient $g_2$ of the expansion \eqref{eq:ExpansionGEnonInfo}. This is the case for the correlation between two bits $x_i^{\mu},x^{\mu}_j$ taken from the same pattern. One has:
\begin{multline}
	\label{eq:CorrBitsPattern}
	\mathds{E}_{\boldsymbol{x}^{\mu}}x^{\mu}_i\,x^{\mu}_j\Theta\left(y^{\mu}\varphi\left(\frac{1}{\sqrt{N}}\sum_ix^{\mu}_iw^*_i\right)\right)=\\=\int\frac{d\hat{h}^{\mu}dh^{\mu}}{2\pi}\mathrm{e}^{i\hat{h}^{\mu}h^{\mu}}\Theta\left(y^{\mu}\varphi(h^{\mu})\right)\left[\prod_{k\neq i,j}\mathds{E}_{x^{\mu}_k}\mathrm{e}^{-i\hat{h}^{\mu}\frac{x^{\mu}_kw^*_k}{\sqrt{N}}}\right]\left[\mathds{E}_{x^{\mu}_i}x^{\mu}_i\mathrm{e}^{-i\hat{h}^{\mu}\frac{x^{\mu}_iw^*_i}{\sqrt{N}}}\right]\left[\mathds{E}_{x^{\mu}_j}x^{\mu}_j\mathrm{e}^{-i\hat{h}^{\mu}\frac{x^{\mu}_jw^*_j}{\sqrt{N}}}\right]=\\=\frac{w_i^*w_j^*}{N}\int Dh\,(h^2-1)\,\Theta\left(y^{\mu}\varphi(h)\right)=\frac{w_i^{*}w_j^{*}}{N}H_2(y^{\mu})\,,
\end{multline}
where $H_2$ is the same function defined in Eq.~\eqref{eq:definitionH0H1H2}. Note that if $\varphi$ is odd, then by symmetry,
\begin{equation}
	\int Dh\,(h^2-1)\,\Theta(\varphi(h))=\int Dh\, (h^2-1)\,\frac{1}{2}\Big(1+\text{sgn}(\varphi(h))\Big)=0\,.
\end{equation}
In general, choosing the activation in such a way that $H_2=0$, it follows that the average correlation between two bits does not carry information about the planted signal. Also, if $H_1=H_2=0$, then the non-informative fixed point is a stable fixed point of the state evolution of AMP for each $\alpha=O(1)$.

\paragraph*{Acknowledgements.}
E.M.M. acknowledges the MUR-Prin 2022 funding Prot. 20229T9EAT, financed by the European Union (Next Generation EU). This paper is supported by PNRR-PE-AI FAIR project funded by the NextGeneration EU program. N.I.S. was supported by European Research Council (ERC) under the EU’s Horizon 2020 research and innovation programme (Grant agreement No. 101019547). A.R. supported by European Research Council (ERC) under the EU’s Horizon 2020 research and innovation programme (Grant agreement No. 101019547) and Cariplo CRYPTONOMEX grant.

\newpage

\bibliographystyle{unsrt}
\bibliography{biblio}

\newpage

\appendix

\section{Computation of the clonated partition function}
\label{app:firstmoment}
\subsection{Annealed Computation}
In the following we use $a = 1\,, \dots \,, m$ to denote different solutions. We use the notation $Dx\equiv dx\, \mathrm{e}^{-x^2/2}/\sqrt{2\pi}$.
We have:
\begin{equation}
	\mathbb{E} \mathcal{N}_m (q)=\frac{1}{2^N}\mathds{E}_A\sum_{\boldsymbol{w}_0\cdots \boldsymbol{w}_m}\prod_{\mu}\prod_a\Theta\left(\varphi\left(\frac{\boldsymbol{x}^{\mu}\cdot\boldsymbol{w}_0}{\sqrt{N}}\right)\,\varphi\left(\frac{\boldsymbol{x}^{\mu}\cdot\boldsymbol{w}_a}{\sqrt{N}}\right)\right)\,\prod_{a<b}\delta\left(q-\frac{\boldsymbol{w}_a\cdot\boldsymbol{w}_b}{N}\right)\,.
\end{equation}
Given a set of configurations $\boldsymbol{w_0},\boldsymbol{w_1},\dots,\boldsymbol{w_m}$, the quantities $\frac{\boldsymbol{x}^{\mu}\cdot \boldsymbol{w_a}}{\sqrt{N}}$ are correlated Gaussian variables.
\begin{equation}
	\label{eq::qab}
	\mathds{E}\left[\frac{\boldsymbol{x}^{\mu}\cdot \boldsymbol{w_a}}{\sqrt{N}}\frac{\boldsymbol{x}^{\nu}\cdot \boldsymbol{w_b}}{\sqrt{N}}\right]=\delta_{\mu\nu}\,q_{ab}\,,
\end{equation}
where if $a,b>0$, $q_{ab}=q$. We call $p_a=q_{0a}$ the overlaps between the teacher and the students. At this point we insert the identity: 
\begin{equation}
	1=\int \prod_adp_a\,\delta\left(p_a-\frac{\boldsymbol{w_0}\cdot\boldsymbol{w_a}}{N}\right)\,.
\end{equation}
In this way we get
\begin{equation}
	\begin{split}
		\mathds{E}\mathcal{N}_m(q)&=\int \prod_adp_a\,\left(\int D_{\Sigma}\boldsymbol{u}\prod_{a=1}^m\Theta(\varphi(u_0)\varphi(u_a))\right)^P\,\frac{1}{2^N}\sum_{w_0\cdots w_m}\,\prod_{a<b}\delta\left(q-\frac{\boldsymbol{w}_a\cdot\boldsymbol{w}_b}{N}\right)\,\prod_a\delta\left(p_a-\frac{\boldsymbol{w}_0\cdot\boldsymbol{w}_a}{N}\right)=\\&=\int \prod_a dp_a\, \mathrm{e}^{mN\,\phi^{\text{ann}}_{m}(q,p_1,\dots,p_m)}\,,
	\end{split}
\end{equation}
where we defined 
\begin{equation}
	\phi^{\text{ann}}_{m}(q,p_1,\dots,p_m)=G_S^{(m)}(q,p_1,\dots,p_m)+\alpha\,G_E^{(m)}(q,p_1,\dots,p_m)\,,
\end{equation}
\begin{equation}
	\label{eq:APP_Gs}
	G_S^{(m)}(q,p_1,\dots,p_m)= \frac{1}{mN}\log{\frac{1}{2^N}\sum_{\boldsymbol{w}_0\cdots \boldsymbol{w}_m}\prod_{a<b}\delta\left(q-\frac{\boldsymbol{w}_a\cdot\boldsymbol{w}_b}{N}\right)\,\prod_a\delta\left(p_a-\frac{\boldsymbol{w}_0\cdot\boldsymbol{w}_a}{N}\right)}\,,
\end{equation}
\begin{equation}
	\label{eq:GenFormGE}
	G_E^{(m)}=\frac{1}{m}\log{\int D_{\Sigma}\boldsymbol{u}\prod_{a=1}^m\Theta(\varphi(u_0)\varphi(u_a))}\,.
\end{equation}
In the large $N$ limit, the annealed free entropy $\frac{1}{N}\log{\mathds{E}\mathcal{N}_m(q)}$ can be computed by steepest descent:
\begin{equation}
	\lim_{N\rightarrow\infty}\frac{1}{N}\log{\mathds{E}\mathcal{N}_m(q)}=\max_{p_1,\dots,p_m}\phi(q,p_1,\dots,p_m)\,.
\end{equation}
In the next sections, we compute the energetic and entropic terms within the symmetric ansatz which assumes that all the clones are equivalent, namely $p_a=p\,\, \forall a$. \\
Let us compute the entropic term $G_S^{(m)}$. Using an integral representation of the Dirac delta function, we have:
\begin{multline}
	G_S^{(m)}(q,p_1,\dots,p_m) =\\ =\frac{1}{mN}\int \prod_a\frac{d\hat{p}_a}{2\pi/N}\prod_{a<b} \frac{d\hat{q}_b}{2\pi/N}\,\mathrm{e}^{-Nq\sum_{a<b}\hat{q}_{ab}-Np\sum_a\hat{p}_a}\left(\frac{1}{2}\sum_{w_0\cdots w_m}\mathrm{e}^{\sum_{a<b}\hat{q}_{ab}w_aw_b+w_0\sum_a \hat{p}_aw_a}\right)^N
	\,.
\end{multline}
The symmetric ansatz corresponds to take:
\begin{equation}
	p^a=p\,,\quad \hat{p}^a=\hat{p}\,,
\end{equation}
\begin{equation}
	\hat{q}^{ab}=(1-\delta_{ab})\,\hat{q}\,.
\end{equation}
In this way, by saddle point approximation we get:
\begin{equation}
	G_S^{(m)}(p,q)=\min_{\hat{q},\hat{p}}G_S^{(m)}(q,p,\hat{q},\hat{p})+o_N(1)\,,
\end{equation}
where,
\begin{equation}
	\label{eq:GSaux}
	\begin{split}
		G_S^{(m)}(q,p,\hat{q},\hat{p})&=-\frac{(m-1)}{2}q\,\hat{q}-p\,\hat{p}-\frac{\hat{q}}{2}+\frac{1}{m}\log{\int Dx\bigg[2\cosh{\left(x\sqrt{\hat{q}}+\hat{p}\right)}\bigg]^m}\\&= -\frac{(m-1)}{2}q\,\hat{q}-p\,\hat{p}-\frac{\hat{q}}{2}+\frac{1}{m}\log{\left(\sum_{k=0}^{m}\binom{m}{k}\mathrm{e}^{(2k-m)\hat{p}+\frac{\hat{q}}{2}(2k-m)^2}\right)}\,.
	\end{split}
\end{equation}
The last identity in \cref{eq:GSaux} follows noting that, 
\begin{equation}
	\int Dx \left(2\cosh{\left(\sqrt{\hat{q}}x+\hat{p}\right)}\right)^m=\sum_{k=0}^{m}\binom{m}{k}\int D x \,\mathrm{e}^{k(\sqrt{\hat{q}}x+\hat{p})-(m-k)(\sqrt{\hat{q}}x+\hat{p})}=\sum_{k=0}^{m}\binom{m}{k}\mathrm{e}^{(2k-m)\,\hat{p}+\frac{\hat{q}}{2}\,(2k-m)^2}\,.
\end{equation}
At this point, let us compute the energetic term $G_{E}^{(m)}$. Using the Gaussian identity,
\begin{equation}
	\frac{\mathrm{e}^{-\frac{1}{2}\boldsymbol{u}^T\Sigma^{-1}\boldsymbol{u}}}{\sqrt{(2\pi)^{n}\,\det{\Sigma}}}=\int\prod_{a=1}^n\frac{d\hat{u}_a}{2\pi}\,\mathrm{e}^{-\frac{1}{2}\boldsymbol{\hat{u}}^T\Sigma \boldsymbol{\hat{u}}+i\,\boldsymbol{u}\cdot \boldsymbol{\hat{u}}}\,,
\end{equation}
the energetic term can be rewritten as follows:
\begin{equation}
	G_E^{(m)} =\frac{1}{m}\ln \int \frac{du d\hat{u}}{2\pi}\mathrm{e}^{-\frac{1}{2}\hat{u}^2-iu\hat{u}}
	\int \prod_{a} \frac{d h^{a} d \hat h^{a}}{2\pi} \mathrm{e}^{i \sum_ah^{a} \hat{h}^{a}-\frac{1}{2} \sum_{ab}q^{ab} \hat h^a \hat h^b-\hat{u}\sum_a\hat{h}^ap^a} \prod_{a} \Theta\left[\varphi(u)\,\varphi(h^a)\right]\,.
\end{equation}
Using the symmetric ansatz, we can write,
\begin{equation}
	-\frac{1}{2}\sum_{ab}q^{ab}\hat{h}^a\hat{h}^b=-\frac{q}{2}\left(\sum_a\hat{h}^a\right)^2-\frac{1-q}{2}\sum_a(\hat{h}^a)^2\,.
\end{equation}
and therefore,
\begin{multline}
	\int \prod_{a} \frac{d h^{a} d \hat h^{a}}{2\pi} \mathrm{e}^{i\sum_a h^{a} \hat{h}^{a}-\frac{1}{2} \sum_{ab}q^{ab} \hat h^a \hat h^b-\hat{u}\sum_a\hat{h}^ap^a} \prod_{a} \Theta\left[\varphi(u)\,\varphi(h^a)\right]=\\=\int Dx \,\left[\int\frac{dhd\hat{h}}{2\pi}\mathrm{e}^{i\hat{h}h-\hat{u}\hat{h}p+ix\sqrt{q}\hat{h}-\frac{1-q}{2}(\hat{h})^2} \Theta\left[\varphi(u)\,\varphi(h)\right]\right]^m=\\=\int Dx \,\left[\int\frac{dh}{\sqrt{2\pi(1-q)}}\mathrm{e}^{-\frac{(h+\sqrt{q}x-ip\hat{u})^2}{2(1-q)}} \Theta\left[\varphi(u)\,\varphi(h)\right]\right]^m\,.
\end{multline}
By doing a change of variables $\sqrt{q} x-i\,p\hat{u}\rightarrow x$, and integrating over $\hat{u}$, we get,
\begin{equation}
	G_E^{(m)}=\frac{1}{m}\log{\int Du\, Dx\left[\int Dh\,\Theta\left(\varphi(u)\,\varphi(\sqrt{1-q}\,h+\sqrt{q-p^2}\,x+p\,u)\right)\right]^m}\,.
\end{equation}
Note that we have,
\begin{equation}
	\begin{split}
		\mathrm{e}^{m\,G_E^{(m)}}&=\int Dx\, Du\,\left[\int Dh\,\Theta\left(\varphi(u)\,\varphi(\sqrt{1-q}\,h+\sqrt{q-p^2}\,x+p\,u)\right)\right]^m=\\&=\int Dx\, Du\,\left[\sum_{\tau=\pm 1}\int Dh\,\Theta\left(\tau\varphi(u)\right)\,\Theta\left(\tau\varphi(\sqrt{1-q}\,h+\sqrt{q-p^2}\,x+p\,u)\right)\right]^m=\\&
		=\sum_{\tau=\pm 1}\int Dx \,D u\,\Theta\left(\tau\varphi(u)\right)\, \left[\int Dh\,\Theta\left(\tau\varphi(\sqrt{1-q}\,h+\sqrt{q-p^2}\,x+p\,u)\right)\right]^m\,.
	\end{split}
\end{equation}
At this point, we change variables $\sqrt{q-p^2}\,x+p\,u\rightarrow x$. In this way we get,
\begin{equation}
	\mathrm{e}^{m\,G_E^{(m)}}=\sum_{\tau=\pm 1}\int Dx\, Du\,\Theta\left(\tau\varphi\left(\sqrt{1-\sfrac{p^2}{q}}\,u+\sfrac{p}{\sqrt{q}}\,x\right)\right)\,\left[\int Dh\,\Theta\left(\tau\varphi(\sqrt{1-q}\,h+\sqrt{q}\,x)\right)\right]^m\,.
\end{equation}
At this point it is convenient to define the kernel $I_{\tau}(m,\sigma)$:
\begin{equation}
	\label{eq::Itau}
	\int Dh\,\Theta\left(\tau\varphi(\sigma\,h+m)\right)=I_{\tau}(m,\sigma)\,,
\end{equation}
that satisfies,
\begin{equation}
	\label{eq:normI}
	\sum_{\tau=\pm 1}I_{\tau}(m,\sigma)=1\,.
\end{equation}
In this way we can rewrite,
\begin{equation}
	\mathrm{e}^{m\,G_E^{(m)}}=\sum_{\tau}\int Dx\, I_{\tau}\left(\sfrac{p}{\sqrt{q}}\,x,\sqrt{1-\sfrac{p^2}{q}}\right)\,I_{\tau}\left(\sqrt{q}\,x,\sqrt{1-q}\right)^m\,.
\end{equation}
Note that if $\varphi(\cdot)$ is odd, we have:
\begin{equation}
	I_{\tau}(m,\sigma)=I_{-\tau}(-m,\sigma)\,,
\end{equation} 
\begin{equation}
	\label{eq:GEodd}
	\mathrm{e}^{m\,G_E^{(m)}}=2\int Dx\, I\left(\sfrac{p}{\sqrt{q}}\,x,\sqrt{1-\sfrac{p^2}{q}}\right)\,I\left(\sqrt{q}\,x,\sqrt{1-q}\right)^m\,,
\end{equation}
where we called
\begin{equation}
	\label{eq:KerneloddDef}
	I(m,\sigma)\equiv I_1(m,\sigma)=\int Dh\,\Theta(\varphi(\sigma\, h +m))\,.
\end{equation}
In the next section, we compute the function $I(m,\sigma)$ for different activations.

\subsection{The Truncated Square Wave}
The ``truncated square wave'' is defined by the activation,
\begin{equation}
	\varphi_{K,\gamma}(h)=\prod_{\ell=-K}^K\text{sgn}\left(h+\frac{\ell\gamma}{K}\right)\,,
\end{equation}
therefore function \eqref{eq:KerneloddDef} is given by
\begin{equation}
	\label{eq:defIfunction}
	\begin{split}
		I_{K,\gamma}(m,\sigma) &\equiv \int Dh \, \Theta\left( \varphi(\sigma\, h + m)\right)=\int_{\frac{\gamma - m}{\sigma}}^\infty D h + \sum_{l=0}^{K-1} \int_{\frac{-\gamma + \frac{2l \gamma}{K} - m}{\sigma}}^{\frac{-\gamma + \frac{(2l+1) \gamma}{K} - m}{\sigma} } D h= \\
		&= H\left( \frac{\gamma - m}{\sigma} \right) + \sum_{l=0}^{K-1}  \left( H\left(\frac{-\gamma + \frac{2l \gamma}{K} - m}{\sigma} \right) - H\left( \frac{-\gamma + \frac{(2l+1) \gamma}{K} - m}{\sigma}\right) \right)\,, 
	\end{split}
\end{equation}
where we used the notation,
\begin{equation}
	H(z)\equiv\int_{z}^{\infty}Dx=\frac{1}{2}\,\text{Erfc}\left(\frac{z}{\sqrt{2}}\right)\,.
\end{equation}
Note that,
\begin{equation}
	I_{K,\gamma}(0,1)=\int Dh\, \Theta\left(\varphi(h)\right)=\frac{1}{2}\,.
\end{equation}

\subsection{The Square Wave}
By taking the limit $\gamma,K\rightarrow \infty$ at fixed ratio $\gamma/K=\delta$, the GRW becomes a square wave with period $T=2\delta$. For this reason it is natural to project the activation on the Fourier basis. We have,
\begin{equation}
	\varphi(x)=a_0+\sum_{n=1}^{\infty}a_n\,\cos{\frac{2\pi nx}{T}}+\sum_{n=1}^{\infty}b_n\,\sin{\frac{2\pi nx}{T}}\,.
\end{equation}
The terms $a_n$ are zero because $\varphi$ is odd, while,
\begin{equation}
	b_n=\frac{2}{T}\int_0^Tdx\,\varphi(x)\,\sin{\frac{2\pi nx}{T}}=\frac{1}{\delta}\int_{\delta}^{2\delta}dx\,\sin{\frac{\pi n x}{\delta}}-\frac{1}{\delta}\int_0^{\delta}dx\,\sin{\frac{\pi n x}{\delta}}=\frac{4\cos{(n\pi)}\,\sin^2{(\frac{n\pi}{2}})}{n\pi},\quad n\geq 1\,.
\end{equation}
Therefore,
\begin{equation}
	\varphi(h)=-\frac{4}{\pi}\sum_{n=0}^{\infty}\frac{1}{2n+1}\sin{\left(\frac{\pi}{\delta}(2n+1)\,h\right)}\,.
\end{equation}
Using this representation in the kernel $I_{\delta}(m,\sigma)$ defining the energetic term of the free entropy, we get,
\begin{equation}
	\label{eq:KernelSWP}
	\begin{split}
		I_{\delta}(m,\sigma)=\int Dh\, \Theta\left(\varphi(\sigma\,h+m)\right)&=\frac{1}{2}+\frac{1}{2}\int Dh\,\varphi(\sigma\,h+m)=\\&=\frac{1}{2}-\frac{2}{\pi}\sum_{n=0}^{\infty}\frac{\mathrm{e}^{-\,\pi^2\frac{\sigma^2}{\delta^2}\,\frac{(2n+1)^2}{2}}}{2n+1}\sin{\left((2n+1)\frac{\pi}{\delta}m\right)}\,.
	\end{split}
\end{equation}
Using Eq.~\eqref{eq:GEodd} we get,
\begin{equation}
	\mathrm{e}^{m\,G_E(q,p)}=\int Dx\,I_{\delta}\left(\sqrt{q}\,x,\sqrt{1-q}\right)^m-\frac{4}{\pi}\sum_{n=0}^{\infty}\frac{\mathrm{e}^{-\frac{\pi^2(2n+1)^2}{2\delta^2}\frac{q-p^2}{q}}}{2n+1}\int Dx\,\sin\left(\frac{(2n+1)\,\pi\, p\, x}{\sqrt{q}\,\delta}\right)\,I_{\delta}\left(\sqrt{q}\,x,\sqrt{1-q}\right)^m\,.
\end{equation}

\section{Cloning in different instances of disorder}
\label{app:eOGP}

In a variation of the $m$-OGP, called ensemble OGP (e-OGP), the clones are allowed to live in different instances of the disorder. This variation of OGP is used to construct a refutation to stable algorithms. 

Consider for simplicity the storage case with $m=2$. We denote by $\boldsymbol{\xi}^{\mu,1}$ and $\boldsymbol{\xi}^{\mu,2}$ the realization of pattern $\mu$ in the first and second instance, respectively. We consider:
\begin{equation}
	\label{eq:defCorr}
	\mathds{E}\,\xi_i^{\mu,1} = \mathds{E}\,\tilde{\xi}_i^{\mu,2}=0,\quad \mathds{E}\,\big(\xi_i^{\mu,1}\big)^2 = \mathds{E}\,\big(\tilde{\xi}_i^{\mu,2}\big)^2=1,\quad \mathds{E}\,\xi_i^{\mu,1}\tilde{\xi}_j^{\nu,2} = \delta_{ij}\delta_{\mu\nu}\rho,
\end{equation}
for some $\rho\in [0,1]$, where $\mathds{E}$ is the expectation over the patterns. The energetic term of the clonated free entropy for $m=2$ reads:
\begin{equation}
	G_{E}^{(2)}=\frac{1}{2}\,\log{\int D_{\Sigma}\boldsymbol{u}}\,\Theta(\varphi(u_1))\,\,\Theta(\varphi(u_2)),
\end{equation}
where the elements of the covariance matrix $\Sigma$ are given by:
\begin{equation}
	\Sigma_{ab}=\delta_{ab}+\rho\,q\,(1-\delta_{ab}).
\end{equation}
Note that the single-instance case that we analyzed in the paper is obtained taking $\rho\to 1$. We call $I(\rho\,q)=\exp{(2\,G_{E}^{(2)})}$, and $r=\rho\,q$. Let \(s(z) = \mathrm{sign}(\varphi(z)) \in\{\pm1\}\) denote the zero-mean square-wave sign so that \(\Theta(\varphi(z))=\tfrac12(1+s(z))\). We define
\[
I(r)=\mathbb{E}\!\left[\Theta(\varphi(u_1))\,\Theta(\varphi(u_2))\right]
=\frac{1}{4}\Big(1+C(r)\Big),\qquad
C(r):=\mathbb{E}[s(u_1)\,s(u_2)].
\]
Every square-integrable function \(f\) of a standard Gaussian variable admits an expansion in the orthonormal Hermite basis,
\[
f(z)=\sum_{n\ge0} a_n\,h_n(z),\qquad
h_n(z):=\frac{H_n(z)}{\sqrt{n!}},\qquad
a_n=\mathbb{E}\big[f(Z)\,h_n(Z)\big],
\]
where \(H_n(z)\) are the (probabilists') Hermite polynomials and \(Z\sim\mathcal{N}(0,1)\). For the odd function \(s(z)\), all even coefficients vanish, \(a_{2k}=0\). Writing \(s(u_a)=\sum_{n\ge0} a_n\,h_n(u_a)\) for \(a\in\{1,2\}\) and using the standard identity for correlated Gaussians
\[
\mathbb{E}\!\left[h_n(u_1)\,h_m(u_2)\right]=\delta_{nm}\,r^{\,n},
\]
we obtain
\begin{equation*}
	\begin{split}
		C(r)&=\mathbb{E}[s(u_1)s(u_2)]
		=\sum_{n,m} a_n a_m\,\mathbb{E}[h_n(u_1)h_m(u_2)]
		=\sum_{n\ge0} a_n^2\,r^{\,n}=\sum_{k\ge0} a_{2k+1}^2\,r^{\,2k+1}.
	\end{split}
\end{equation*}
Differentiating with respect to \(r\) gives
\[
C'(r)=\sum_{k\ge0}(2k{+}1)\,a_{2k+1}^2\,r^{\,2k}\ge0,
\]
with strict inequality for nontrivial \(s\). This means that $G_E^{(2)}(\rho\,q)$ is a strictly increasing function of $\rho\in[0,1]$, implying that 
\begin{equation}
	\label{eq:upbouneOGP}
	\alpha_{2OGP}^{\text{ann}}(\rho)\leq \alpha_{2OGP}^{\text{ann}}(1),
\end{equation}
where $\alpha_{2\mathrm{OGP}}^{\text{ann}}(\rho)$ denotes the annealed 2-OGP threshold computed with the two clones living in correlated instances, and $\alpha_{2\mathrm{OGP}}^{\text{ann}}(1)$ the corresponding threshold when both clones belong to the same instance. In other words, the presence of an OGP within the same instance implies its presence across two instances with correlation defined in Eq.~\eqref{eq:defCorr}. In this sense, the single-instance setting represents a worst-case scenario, which justifies our focus on this case. A similar argument can be done for $m>2$ assuming uniform correlation between the different $m$ instances. 

\section{The Replica-Symmetric Free Entropy of One Clone}
\label{app:RSm0existence}
By taking the limit $m\rightarrow 0$, optimizing over $q$ and $p$ is equivalent to computing the free entropy in a replica-symmetric ansatz. For the entropic part we get,
\begin{equation}
	G_S=\frac{1}{2}q\,\hat{q}-p\,\hat{p}-\frac{\hat{q}}{2}+\int Dx\,\log{\left(2\cosh{\left(x\sqrt{\hat{q}}+\hat{p}\right)}\right)}\,.
\end{equation}
Using \eqref{eq:normI} we get,
\begin{equation}
	G_E=\sum_{\tau}\int Dx\, I_{\tau}\left(\sfrac{p}{\sqrt{q}}\,x,\sqrt{1-\sfrac{p^2}{q}}\right)\,\log{I_{\tau}\left(\sqrt{q}\,x,\sqrt{1-q}\right)}\,,
\end{equation}
that for odd activations becomes,
\begin{equation}
	G_E=2\int Dx\, I\left(\sfrac{p}{\sqrt{q}}\,x,\sqrt{1-\sfrac{p^2}{q}}\right)\,\log{I\left(\sqrt{q}\,x,\sqrt{1-q}\right)}\,,
\end{equation}
where $I_{1}(\cdot,\cdot)\equiv I(\cdot,\cdot)$. Note that sending $p,q\rightarrow 0$ one recovers the annealed free entropy $\phi=(1-\alpha)\log{2}$ for the existence of solutions.

It can be useful to write the saddle-point equations for the entropic part, which read:
\begin{equation}
	q=1-\int Dx\,\frac{x}{\sqrt{\hat{q}}}\,\tanh{\left(\sqrt{\hat{q}}\,x+\hat{p}\right)}=\int Dx\, \tanh{\left(\sqrt{\hat{q}}\,x+\hat{p}\right)}^2\,,
\end{equation}
\begin{equation}
	p=\int Dx\,\tanh{\Big(\sqrt{\hat{q}}\,x+\hat{p}\Big)}\,.
\end{equation}
Under the Nishimori condition $q=p$ and $\hat{q}=\hat{p}$.

\subsection{Annealed Upper Bound on the Storage Capacity}
\label{app:annealedStoragecap}
We have to compute the expected value:
\begin{equation}
	\mathds{E}_{\mathcal{D}}\mathcal{N}_1=\mathds{E}_{\mathcal{D}}\sum_{\boldsymbol{w}}\prod_{\mu=1}^P\Theta\left[ y^{\mu} \, \varphi\left(\frac{1}{\sqrt{N}}\boldsymbol w\cdot\boldsymbol x^{\mu} \right)\right]=2^N\prod_{\mu}\int Dh\,\Theta\left(y^{\mu}\,\varphi(h)\right).
\end{equation}
In the SWP case, and in general for all asymmetric activations, one has,
\begin{equation}
	\int Dh\,\Theta\left(y^{\mu}\,\varphi(h)\right)=\frac{1}{2}\,,
\end{equation}
independently of $y^{\mu}$. Therefore,
\begin{equation}
	\phi^{\text{ann}}_1=(1-\alpha)\log{2}\,.
\end{equation}

\subsection{Annealed Upper Bound on the Teacher-Student}
\label{app:annealedTeacherStudent}

Computing the annealed clonated free entropy for $q\rightarrow 1$ (all clones coincide) and $m=1$, one gets the annealed free entropy of the teacher-student problem. This free-entropy reads,
\begin{equation}
	\phi^{\mathrm{ann}}_1(\alpha)=\max_p \{G_S(p)+\alpha\,G_E(p)\}\,,
\end{equation}
where,
\begin{equation}
	G_S(p)=-p\,\tanh^{-1}(p)+\log{\left(\frac{2}{\sqrt{1-p^2}}\right)}\,,
\end{equation}
\begin{equation}
	G_E(p)=\log{\left\{2\int Dx\, I\left(p\,x,\sqrt{1-p^2}\right)\,\Theta(\varphi(x))\right\}}\,.
\end{equation}
The annealed free entropy provides an upper-bound to the teacher threshold, by computing the point $\alpha$ such that $\phi^{\mathrm{ann}}_1(\alpha)=0$. Note that a lower bound for $\alpha_T$ is one.

\section{Satisfiability threshold in the small $\delta$ limit}
\label{app:SecondMoment}
Using a second-moment criterium, it is possible to argue that the storage capacity of the SWP saturates the annealed upper bound in the small $\delta$ limit. Using the Cauchy-Schwartz inequality,
\begin{equation}
	\mathbb{E}\, \mathcal{N}_1\leq \text{Pr}_{\mathcal{D}}\big[\mathcal{N}_1>0\big]^{1/2}\,\left(\mathbb{E}\, \mathcal{N}_1^2\right)^{1/2}\,,
\end{equation}
and therefore, if $\mathbb{E}\, \mathcal{N}_1^2>0$, we have:
\begin{equation}
	\label{eq:2mom1}
	\frac{\left(\mathbb{E} \mathcal{N}_1\right)^2}{\mathbb{E} \mathcal{N}_1^2}\leq \text{Pr}_{\mathcal{D}}\big[\mathcal{N}_1 > 0\big]\,.
\end{equation}
The numerator of the previous expression is simply given in terms of the annealed free entropy of one clone $\phi^{\text{ann}}_1=(1-\alpha)\,\log{2}$:
\begin{equation}
	\mathds{E}\mathcal{N}_1=e^{N(1-\alpha)\log{2}},
\end{equation}
while the denominator is
\begin{equation}
	\label{eq:2momGen}
	\mathds{E}\mathcal{N}_1^2=\int dq\, e^{2N\,\left(G_S^{(2)}(q)+\alpha\,G_E^{(2)}(q)\right)},
\end{equation}
where the expressions for $G_S^{(2)}$ and $G_E^{(2)}$ are given by
\begin{equation}
	G_S^{(2)}(q)= \lim_{N\rightarrow \infty}\frac{1}{2N}\log{\frac{1}{2^N}\sum_{\boldsymbol{w}_1,\boldsymbol{w}_2}\delta\left(q-\frac{\boldsymbol{w}_1\cdot\boldsymbol{w}_2}{N}\right)},
\end{equation}
\begin{equation}
	G_E^{(2)}=\frac{1}{2}\log{\int Dx\, I\left(\sqrt{q}x,\sqrt{1-q}\right)^2}.
\end{equation}
At this point note that: i) assuming $q < 1$, which can be verified numerically as discussed shortly, we have that as $\delta \to 0$, $G_E^{(2)}(q) \to -\log{2}$; ii) $G_S^{(2)}(q)\leq G_S^{(2)}(0)=\log{2}$. Therefore, for $\delta\rightarrow 0$ we have 
\begin{equation}
	\mathds{E}\mathcal{N}_1^2\leq e^{2\,(1-\alpha)\, N\,\log{2}}.
\end{equation}
Using~\eqref{eq:2mom1} therefore we have that if $\alpha<1$ the probability that an instance is satisfiable is lower-bounded by one. Therefore, since $\alpha=1$ is an upper bound to the critical capacity, this bound is tight in the limit $\delta=0$. 

\begin{figure}[]
	\centering
	\begin{subfigure}{0.49\textwidth}
		\centering
		\includegraphics[width=\textwidth]{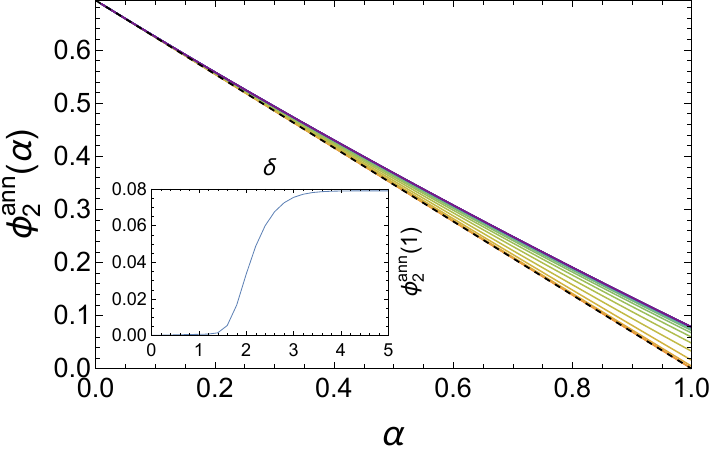}
		\caption{}
		\label{fig:SecondMoment}
	\end{subfigure}
	\hfill
	\begin{subfigure}{0.49\textwidth}
		\centering
		\includegraphics[width=\textwidth]{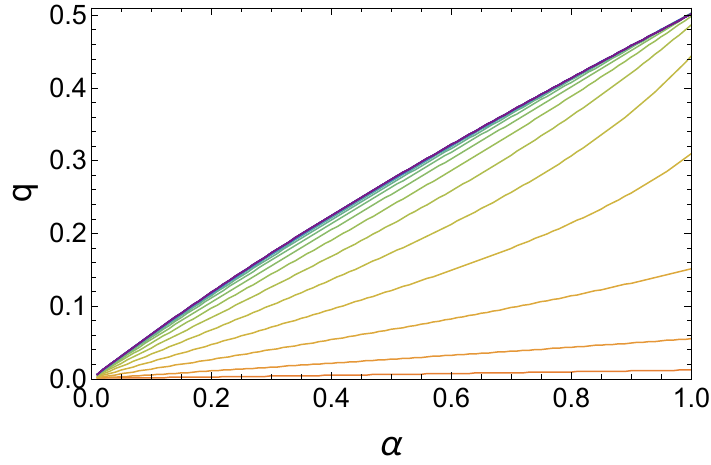}
		\caption{}
		\label{fig:SecondMomentOverlap}
	\end{subfigure}
	\caption{(\emph{a}): Second moment free entropy $\max_{q}\phi^{\text{ann}}_2(q,\alpha)$ as a function of $\alpha$ for different values of $\delta$. From bottom to top $\delta=1.2+n\,0.2$, with $n=0,\dots,19$. The dashed line corresponds to $(1-\alpha)\log{2}$. For $\delta\rightarrow 0$, $(1-\alpha)\log{2}-\max_{q}\phi^{\text{ann}}_2(q,\alpha)\rightarrow 0$ for all values of $0<\alpha<1$. \emph{Inset}: $\max_{q}\phi^{\text{ann}}_2(q,\alpha=1)$ as a function of $\delta$. (\emph{b}): optimal overlap of the second moment free entropy, $\text{argmax}\phi_1^{ann}(q,\alpha)$, as a function of $\alpha$ for different values of $\delta$. From bottom to top $\delta=1.2+n\,0.2$, with $n=0,\dots,19$. Note that reducing $\delta$, also the optimal overlap reduces for all values of $\alpha$.}
\end{figure}

Let us discuss the assumption $q < 1$. For finite $\delta$, the integral \eqref{eq:2momGen} can be computed at the leading order by steepest descent, $\phi_2^{ann}\equiv\lim_{N\rightarrow \infty}\frac{1}{N}\log{\mathds{E}\mathcal{N}_1^2}=\max_q\{G_S^{(2)}(q)+\alpha\,G_S^{(2)}(q)\}$. In Figs.~\ref{fig:SecondMoment} and \ref{fig:SecondMomentOverlap} we show the numerical computation of $\phi_2^{ann}$, and the value of the optimum overlap as a function of $\alpha$, for different values of $\delta$. Observe that the optimum overlap is always strictly smaller than one. In particular, reducing $\delta$, the optimum overlap goes to zero independently from $\alpha$, and $\phi_2^{ann}\rightarrow (1-\alpha)\log{2}$, therefore, the assumption $q<1$ is verified.

\section{Non-monotonicity of the hard--easy transition}~\label{app::non-monotonicity_hard-easy}

While the divergence of the hard--easy transition as $\delta \to 0$ admits a simple interpretation in terms of disorder observables (see Sec.~\ref{sec:ComputationDistinguishers}), the initial \emph{decrease} of the transition when lowering $\delta$ from $\infty$ is more subtle. This feature can be seen directly from the RS formulas by comparing the energetic terms of the free entropies. Indeed, we can decompose the RS potential as
\begin{equation}
	\label{eq:sumsub}
	\Phi_{\text{SWP}}(p,\alpha) \;=\; G_S(p)\;+\;\alpha\,G_E^{\text{SWP}}(p) \;=\; \Phi_{\text{ABP}}(p,\alpha)\;+\;\alpha\,\Delta G_E(p,\delta),
\end{equation}
where $\Phi_{\text{SWP}}$ and $\Phi_{\text{ABP}}$ are the free entropies of the square--wave and asymmetric binary perceptron respectively, on the Nishimori line, and $\Delta G_E(p,\delta)=G_E^{\text{SWP}}(p,\delta)-G_E^{\text{ABP}}(p)$, where $G_E^{\text{SWP}}(p,\delta)$ and $G_E^{\text{ABP}}(p)$ are obtained from equation A.26 by substituting respectively $\varphi(h) = - \mathrm{sign} \left( \sin\left(\frac{\pi h}{\delta}\right)\right)$ and $\varphi(h) = \mathrm{sign}(h)$, and taking $p=q$ (Nishimori condition).

\begin{figure}[h]
	\centering
	\includegraphics[width=0.6\linewidth]{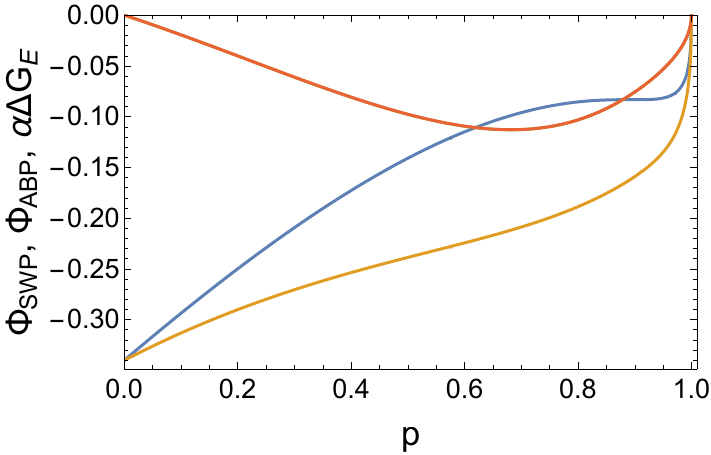}
	\caption{From top left to bottom left: $\alpha\Delta G_E$, $\Phi_{ABP}$ and $\Phi_{SWP}$ computed for $\alpha=\alpha_{ABP}$, $\delta=2.1$. Note that $\Phi_{SWP}$ is monotonously increasing, meaning that AMP messages initialized in a completely uninformed way ($p=0$) would converge to the teacher. This means that $\alpha_{ABP}$ is in the easy phase for SWP with $\delta=2.1$.}
	\label{fig:nonMon}
\end{figure}

For $\alpha=\alpha_{\text{ABP}}$ (the ABP hard--easy transition point), one has $d\Phi_{\text{ABP}}/dp>0$ for all overlaps $p\neq p_{\text{ABP}}$, and $d\Phi_{\text{ABP}}/dp=0$ only at the ABP maximizer $p_{\text{ABP}}\simeq 0.89$ (with $\alpha_{\text{ABP}}\simeq 1.49$, see \cite{gyorgyi1990first}). By analyzing the correction term $\Delta G_E$, one finds that for sufficiently large $\delta$ the total derivative
\[
\frac{d\Phi_{\text{SWP}}}{dp}=\frac{d\Phi_{\text{ABP}}}{dp} + \alpha^{\text{ABP}}\,\frac{d\Delta G_E}{dp}
\]
remains strictly positive for all values of the overlap $p$. In particular, this implies $d\Delta G_E(p^{\text{ABP}},\delta)/dp>0$. The global positivity excludes the appearance of spurious stationary points and guarantees that the SWP hard--easy transition must occur at a strictly smaller value of $\alpha$ than in ABP. 

This reasoning is illustrated in Fig.~\ref{fig:nonMon}, where we plot $\Phi_{\text{SWP}}(p)$, $\Phi_{\text{ABP}}(p)$, and the correction term $\alpha\Delta G_E(p,\delta)$ at $\alpha=\alpha_{\text{ABP}}$ and $\delta=2.1$. We stress that $\Delta G_E(p,\delta)$ is a simple function of $p$ that can be studied without any further optimization. Providing a more geometric interpretation of the behavior of $\Delta G_E$ would be interesting, but we leave this point for future work. 

We note that, in contrast to AMP, the information--theoretic threshold decreases monotonically with $\delta$ and approaches one as $\delta \to 0$, since finer square--wave constraints progressively eliminate other vertices of the hypercube that reproduce the same labels as the teacher at lower values of $\alpha$. In the limit $\delta \to 0$, the teacher becomes information--theoretically identifiable as soon as $\alpha>1$, for instance by exhaustive search. Thus, the best possible algorithm, in terms of the number of examples required to recover the teacher, improves its performance as $\delta$ decreases. A possible explanation of the non--monotonic behavior of AMP is that the algorithm can only exploit a restricted part of the available signal: for large $\delta$ most of the information lies in low modes that AMP can capture, while decreasing $\delta$ gradually shifts weight to higher modes. This may initially improve the effective performance of AMP, mimicking the optimal case, but eventually degrades it once the relevant information is concentrated in modes beyond its reach, thereby producing the observed non--monotonicity.

\section{Number of $m$-Cliques in the Large $m$ Limit}
\label{app:yInfEntropic}
\begin{figure}
	\centering
	\includegraphics[width=0.6\linewidth]{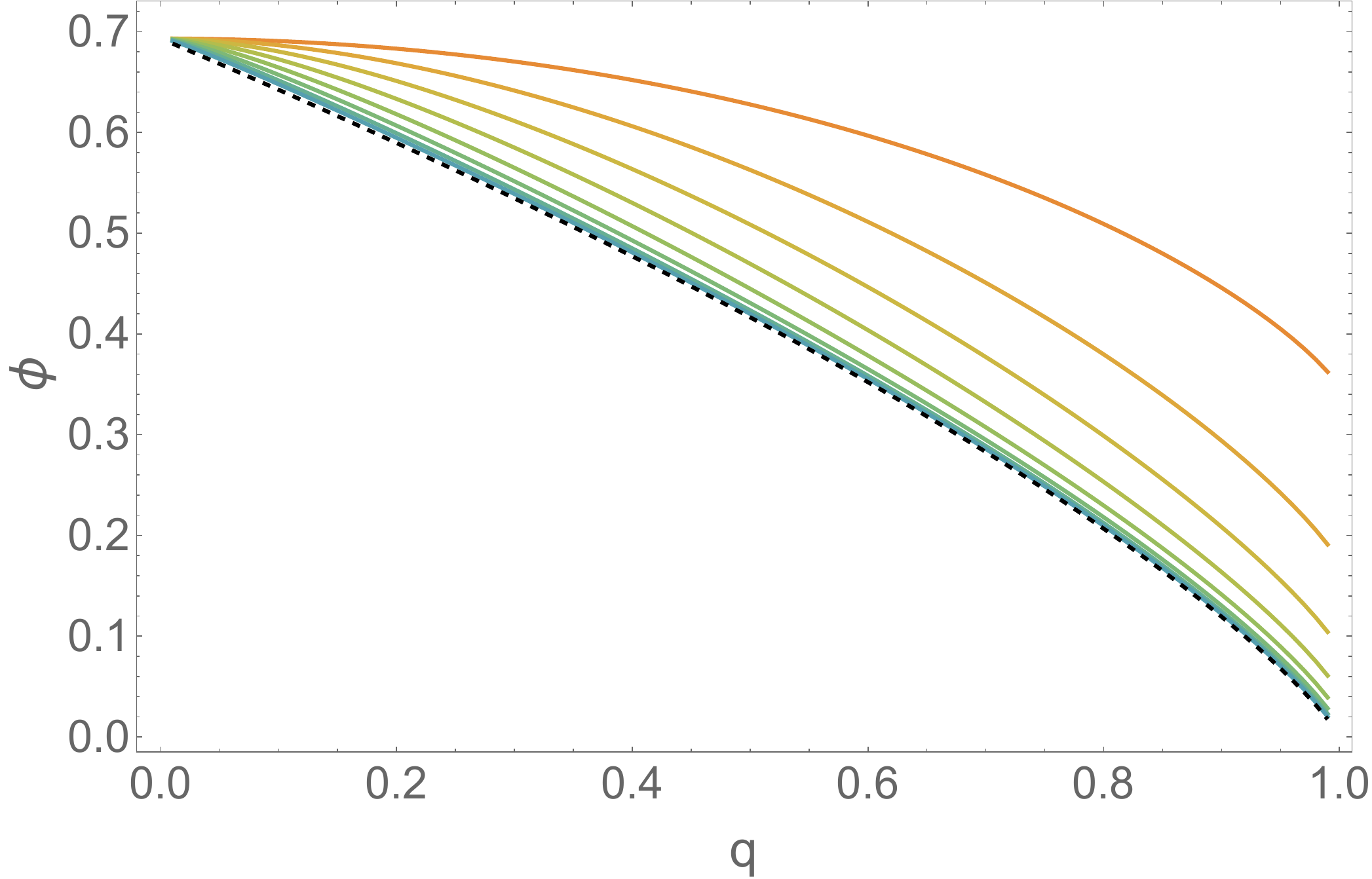}
	\caption{Entropy of $m$ configurations at mutual overlap $q$, and overlap for $m=2^n$. From top to bottom, the continuous lines represent the cases $n=2,\dots,8$, that are obtained solving numerically \eqref{eq:EntropicPart}. The dashed line is obtained taking the limit $m\rightarrow\infty$ (Eq.~ \eqref{eq:EntropicPartinf}). Note that there are huge finite-$m$ corrections.}
	\label{fig:EntropyYConfigvS}
\end{figure}

Let us compute the expected value of the number of $m$-tuples of points of the hypercube having overlap $q$ (cliques). This number is given by the entropic part of the free entropy:
\begin{equation}
	\label{eq:EntropicPart}
	G_S^{(m)}(q)=-\frac{m-1}{2}q\,\hat{q}-\frac{\hat{q}}{2}+\frac{1}{m}\log{\int Dx \left(2\cosh{\left(\sqrt{\hat{q}}\,x\right)}\right)^m}\,.
\end{equation}
In order for the entropy density to be finite, at the fixed point one should have $\hat{q}\sim u/y$. Therefore one finds
\begin{equation}
	\label{eq:EntropicPartinf}
	G_S^{(m)}(q)=\min_{u}\left\{-\frac{1}{2}q\,u+\max_{x}\left\{-\frac{x^2}{2}+\log{\left(2\cosh{\sqrt{u}}\,x\right)}\right\}\right\}\,.
\end{equation}
The associated fixed point equations read,
\begin{equation}
	q=\frac{x\tanh{\left(x\sqrt{u}\right)}}{\sqrt{u}}\,,
\end{equation}
\begin{equation}
	x=\sqrt{u}\tanh{\left(x\sqrt{u}\right)}\,.
\end{equation}
For $q=0$, one has $G_S=\log{2}$. For $q=1$, $G_S=0$. See \cref{fig:EntropyYConfigvS}.

\section{Expansion for small $\delta$}
\label{app:expSmallDelta}

\begin{figure}[]
	\centering
	\begin{subfigure}{0.49\textwidth}
		\centering
		\includegraphics[width=\textwidth]{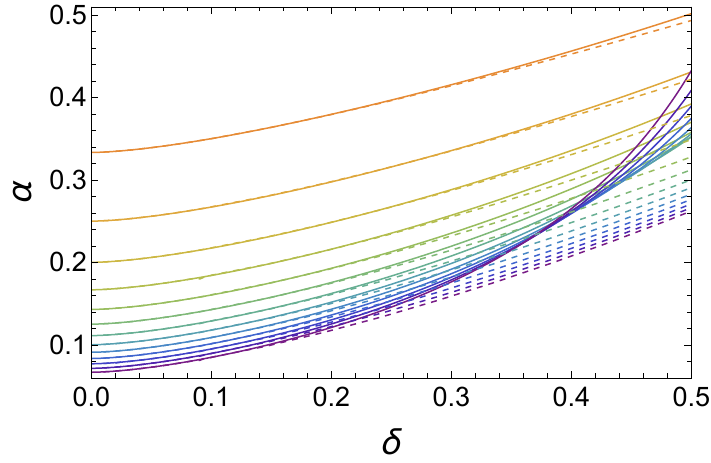}
		\caption{}
		\label{fig:ExpansionOGP}
	\end{subfigure}
	\hfill
	\begin{subfigure}{0.49\textwidth}
		\centering
		\includegraphics[width=\textwidth]{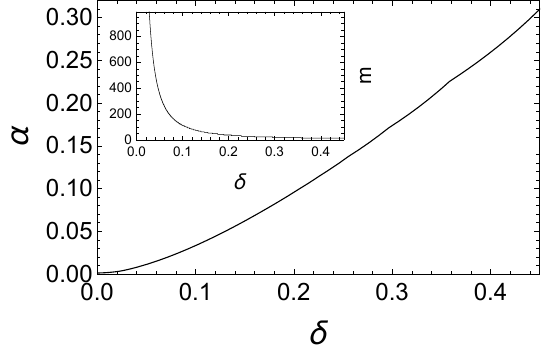}
		\caption{}
		\label{fig:limitcurve}
	\end{subfigure}
	\caption{(\emph{a}): Comparison between the $\alpha^{\text{ann}}_{OGP}(m)$ obtained from the numerical solution of Eqs.~\eqref{eq:conditionsAlphaOGP} (continuous lines) and the expansions around $\delta$ small (dashed lines). From top to bottom $m=3,\dots,15$. (\emph{b}): $\min_m\widetilde{\alpha}^{\text{ann}}_{OGP}(m,\delta)$ as a function of $\delta$. Inset: $\mathrm{argmin}_m\widetilde{\alpha}^{\text{ann}}_{OGP}(m,\delta)$ as a function of $\delta$. For $\delta\to0$ the optimal $m$ diverges. }
\end{figure}

In this section, we study the appearance of an overlap gap in the annealed free entropy in the limit of small $\delta$. It is possible to argue that in this limit, the overlap at which OGP develops tends to one. Indeed, using representation \eqref{eq:Ideltarew}, if one takes the limit $\delta\to\infty$ at fixed $q$, the energetic term tends to a constant, $G_E^{(m)}\to-\log{2}$, and the only dependence on $q$ is on the entropic part. Therefore, the minimum of the free entropy corresponds to the minimum of the entropic term, $G_S^{(m)}(q)$, which is attained at $q=1$, and is equal to $G_S^{(m)}(1)=\frac{1}{m}\log{2}$. Therefore, for $\delta$ small, we can write $q\approx 1-\varepsilon$, and we can expand Eqs.~\eqref{eq:conditionsAlphaOGP} for small $\varepsilon,\delta$. In the following, we assume the scaling $\delta^2=o(\varepsilon)$, and verify the consistency of this assumption at the end of the computation. Let us start from the expression of the energetic term of the free entropy
\begin{equation}
	G_E^{(m)}(q)=\frac{1}{m}\log{\left[\int Dx \left[I\left(\sqrt{q}\,x,\sqrt{1-q}\right)\right]^m\right]},
\end{equation}
where the function $I(\bullet,\bullet)$ is given by:
\begin{equation}
	\label{eq:innerInt}
	I\left(\sqrt{q}\,x,\sqrt{1-q}\right)=\int Dy\, \Theta\left(\varphi\left(\sqrt{1-q}\,y+\sqrt{q}\, x\right)\right).
\end{equation}
If $\delta^2=o(\varepsilon)$, it is convenient to use the Fourier representation of $I$, 
\begin{equation}
	I_\delta\left(\sqrt{q}\, x,\sqrt{1-q}\right)=\frac{1}{2}-\frac{2}{\pi}\sum_{n=0}^{\infty}\frac{\mathrm{e}^{-\,\pi^2\frac{1-q}{\delta^2}\,\frac{(2n+1)^2}{2}}}{2n+1}\sin{\left((2n+1)\frac{\pi}{\delta}\sqrt{q}\,x\right)}.    
\end{equation}
In this regime, all the terms of the series tend to zero, and only the first term determines the behavior of $q$ in the limit $\delta \to 0$. Keeping only the first term one gets:
\begin{equation}
	\label{eq:expandedGe}
	G_E^{(m)}(\varepsilon)\approx -\log{2}+\frac{4}{\pi^2}(m-1)\,e^{-\pi^2\frac{\varepsilon}{\delta^2}}.
\end{equation}
Expanding for small $\varepsilon$ the entropic term (see Eq.~\eqref{eq:GSaux}), one gets 
\begin{equation}
	\label{eq:expandedGs}
	G^{(m)}_S(\varepsilon)\approx\frac{\log{2}}{m}+\varepsilon\,\left(c_m+d_m\log{\frac{2}{\varepsilon}}\right),
\end{equation}
where 
\begin{equation}
	c_2=d_2=1,\quad c_m=\frac{1}{4}\,\,\,\,\,d_m=\frac{1}{2}\quad m>2. 
\end{equation}
Using \eqref{eq:expandedGe} and \eqref{eq:expandedGs} into Eqs.~\ref{eq:conditionsAlphaOGP}, we find that 
\begin{equation}
	\varepsilon(\delta)\approx\delta^2\,y_m(\delta),
\end{equation}
where $y_m(\delta)$ is the solution of: 
\begin{equation}
	\log{y}+A_m(\delta)\,e^{-\pi^2\,y}=B_m(\delta).
\end{equation}
The coefficients $A_m(\delta)$ and $B_m(\delta)$ depend on $m$ and $\delta$. For $m=2$:
\begin{equation}
	A_2(\delta)=\frac{2}{\delta^2},\quad B_2(\delta)=\log{\frac{2}{\delta^2}},
\end{equation}
and for $m>2$
\begin{equation}
	A_m(\delta)=\frac{16}{\delta^2}\frac{m-1}{m},\quad B_m(\delta)=\log{\frac{4}{\delta^2}}.
\end{equation}
Note that $y_m(\delta)$ diverges for $\delta\to 0$, but $\delta^2\,y_m(\delta)\to 0$. Therefore, $\delta^2=o(\varepsilon)$ consistently with the scaling we chose for the overlap. Given $\varepsilon(\delta)$, the critical constraint density is simply given by $\widetilde{\alpha}^{\text{ann}}_{OGP}(m,\delta)\approx-G^{(m)}_S(\varepsilon)/G^{(m)}_E(\varepsilon)$. For small $\delta$:
\begin{equation}
	\alpha^{\text{ann}}_{OGP}(m)\approx \frac{1}{m}+a_m\,\varepsilon\log{\frac{2}{\varepsilon}},
\end{equation}
where 
\begin{equation}
	a_2=\frac{1}{\log{2}},\quad a_{m}=\frac{1}{2\log{2}}\quad \,m>2.
\end{equation}
In~\cref{fig:ExpansionOGP} we compare the small $\delta$ expansion of the annealed OGP thresholds $\widetilde{\alpha}^{\text{ann}}_{OGP}(m,\delta)\approx -G^{(m)}_S(\varepsilon)/G^{(m)}_E(\varepsilon)$, with the estimates $\alpha^{\text{ann}}_{OGP}(m,\delta)$ obtained from the numerical solution of \eqref{eq:conditionsAlphaOGP}. Note that $\widetilde{\alpha}^{\text{ann}}_{OGP}(m,\delta)\geq \alpha^{\text{ann}}_{OGP}(m,\delta)$, therefore we can use $\min_m \widetilde{\alpha}^{\text{ann}}_{OGP}(m,\delta)$ as an upper bound to the OGP threshold. In~\cref{fig:limitcurve} we represent the curve $\min_m \widetilde{\alpha}^{\text{ann}}_{OGP}(m,\delta)$.  

\section{Estimating the $m$-OGP using a Replica-Symmetric Ansatz}
\label{sec:RS_mOGP}

As shown in the main text, the quenched clonated free energy~\eqref{eq:avfreeEntr} is only upper bounded by the annealed entropy that we have considered so far. The annealed computation of the $m$-OGP transition may give rise to inconsistencies, as also noted in the main text. We consider here a more refined computation of the $m$-OGP transition, which is based on a replica-symmetric (RS) ansatz, introduced after using the replica trick,
\begin{equation}
	\frac{1}{m} \mathbb{E}_\mathcal{D} \, \log \mathcal{N}_m(q_1) = \lim_{n\to 0} \frac{\ln \mathbb{E}_{\mathcal{D}} \mathcal{N}^{n/m}_m(q_1)}{n}\,.
\end{equation}
Note that we introduced $n/m$ replicas of the clones, so now we have a total number $n$ of weights $\boldsymbol{w}^a$, $a=1,\dots,n$. After performing the average over inputs, one gets the overlap matrix
\begin{equation}
	q^{ab} \equiv \frac{1}{N} \sum_{i=1}^N w_i^a w_i^b \,.
\end{equation}
Because of the delta function constraint in $\mathcal{N}_m(q_1)$ that enforces an overlap $q_1$ between the clones, the $n/m$ blocks of size $m \times m$ along the diagonal of $q^{ab}$ are fixed to $q$ if $a \ne b$ (and 1 otherwise). The RS ansatz corresponds to suppose that outside those blocks the overlap matrix is parameterized by just a scalar overlap that we will call $q_0$. 

\begin{figure}
	\centering
	\includegraphics[width=0.65\linewidth]{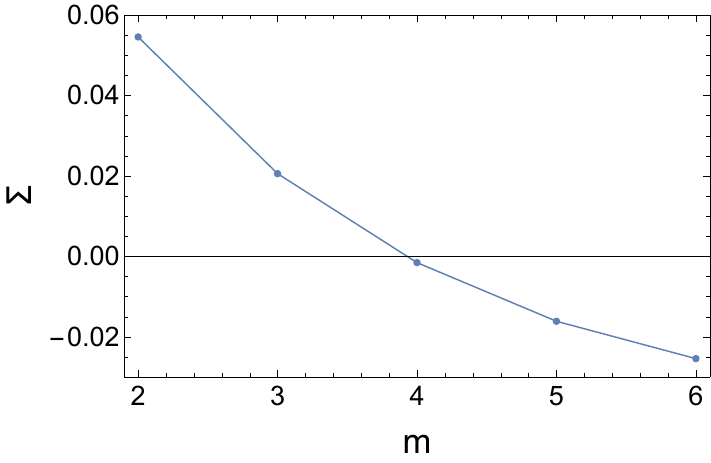}
	\caption{Complexity $\Sigma$ as a function of $m$ for $\delta=1.5$. The negative sign of the complexity for $m\ge 4$ signals that the RS ansatz gives non-physical results in this regime.}
	\label{fig:RSComplexity}
\end{figure}

The expert reader may recognize that imposing an RS ansatz on the clonated free entropy~\eqref{eq:avfreeEntr} is effectively equivalent to imposing a 1-step Replica Symmetry Breaking ansatz (1RSB) on the equilibrium free entropy. The key difference, however, is that in computing~\eqref{eq:avfreeEntr}, one must fix both the Parisi block size $m$ and the overlap $q_1$, rather than optimizing over them as in standard equilibrium calculation. Moreover, $m$ need not to be in the physical range $0< m < 1$. 
In the limit $n\to 0$, one gets,
\begin{subequations}
	\begin{align}
		\phi_m^{\mathrm{rs}}(q_1) &= \text{extr}_{q_0, \hat q_1 \hat q_0} \left[ \mathcal{G}_S(q_1, q_0, \hat q_1, \hat q_0) + \alpha \mathcal{G}_E(q_1, q_0) \right] \\
		\mathcal{G}_S &= - \frac{\hat q_1}{2} (1 - q_1) + \frac{m}{2} \left(q_0 \hat q_0 - q_1 \hat q_1 \right)  - p \hat p 
		+ \frac{1}{m} \int Du \ln \int Dv \, \left(2 \cosh\left(\sqrt{\hat q_0} u + \sqrt{\hat q_1 - \hat q_0} v + \hat p \right) \right)^m\\
		\mathcal{G}_E &= \frac{1}{m} \int Du Dx \ln \int Dy \left[ \int Dh \, \Theta\left( \varphi(u) \varphi(\sqrt{1-q_1} h + p u + \sqrt{q_0 - p^2} x + \sqrt{q_1 - q_0} y) \right) \right]^m\,.
	\end{align}
\end{subequations}
Similarly to the annealed case, the energetic term can be manipulated to get,
\begin{equation}
	\begin{split}
		G_E &= \frac{1}{m} \int Du Dx \ln \int Dy \left[ \int Dh \, \Theta\left( \varphi(u) \varphi(\sqrt{1-q_1} h + p u + \sqrt{q_0 - p^2} x + \sqrt{q_1 - q_0} y) \right) \right]^m \\
		&= \frac{1}{m} \sum_{\tau = \pm 1} \int Du Dx \, \Theta\left( \tau \varphi(u)\right) \ln \int Dy \left[ \int Dh \, \Theta\left( \tau \varphi \left(\sqrt{1-q_1} h + p u + \sqrt{q_0 - p^2} x + \sqrt{q_1 - q_0} y \right) \right) \right]^m \\
		&= \frac{1}{m} \sum_{\tau = \pm 1} \int Du Dx \, \Theta\left( \tau \varphi\left( \frac{p}{\sqrt{q_0}} x + \sqrt{1 - \frac{p^2}{q_0}} u \right)\right)\times\\ &\hspace{6cm} \times  \ln \int Dy \left[ \int Dh \, \Theta\left( \tau \varphi \left(\sqrt{1-q_1} h + \sqrt{q_0} x + \sqrt{q_1 - q_0} y \right) \right) \right]^m\,.
	\end{split}
\end{equation}
i.e.,
\begin{equation}
	\mathcal{G}_E = \frac{1}{m} \sum_{\tau = \pm 1} \int Dx \, I_{\tau}\left(\frac{p}{\sqrt{q_0}} x, \sqrt{1 - \frac{p^2}{q_0}} \right) \ln \int Dy \, I_\tau^m\left( \sqrt{q_0} x + \sqrt{q_1 - q_0} y, \sqrt{1-q_1} \right)\,,
\end{equation}
where $I_\tau$ is the same quantity defined in~\eqref{eq::Itau}. 
Moreover in the case of an odd activation,
\begin{equation}
	\mathcal{G}_E = \frac{2}{m} \int Dx \, I\left(\frac{p}{\sqrt{q_0}} x, \sqrt{1 - \frac{p^2}{q_0}} \right) \ln \int Dy \, I^m\left( \sqrt{q_0} x + \sqrt{q_1 - q_0} y, \sqrt{1-q_1} \right)\,.
\end{equation}
with $I$ defined in~\eqref{eq:KerneloddDef}. Similarly to the annealed case, the RS estimation of the $m$-OGP transition $\alpha^{\mathrm{rs}}_{\mathrm{ogp}}(m)$ can be found by searching for the first value of $\alpha$ where $\phi_m^{\mathrm{rs}}$ has a minimum at $q_1^\star$ with zero entropy, i.e.
\begin{subequations}
	\begin{align}
		\phi_m^{\mathrm{rs}}(q_1^\star) &= 0\, \\
		\left. \frac{\partial \phi_m^{\mathrm{rs}}(q_1)} {\partial q_1} \right|_{q_1 = q_1^\star} &= 0 \,.
	\end{align}
\end{subequations}
Once the $m-$OGP transition is found, one can control whether the solutions are physically admissible by computing the so called complexity or configurational entropy $\Sigma(m)$. It is derived as follows. The cloned partition function, can be written as a sum over pure states each with free entropy $\phi_\alpha$,
\begin{equation}
	\mathcal{N}_m = \sum_\alpha \mathrm{e}^{N m \phi_\alpha} = \int d\phi \sum_\alpha \delta(\phi - \phi_\alpha) \mathrm{e}^{N m \phi} = \int d\phi \, \mathrm{e}^{ N (m\phi + \Sigma(\phi))}\,,
\end{equation}
where we have introduced the complexity or configurational entropy as the log of the number of states $\alpha$ having free entropy $\phi$, i.e.,
\begin{equation}
	\Sigma(\phi) \equiv \frac{1}{N} \ln \sum_\alpha \delta(\phi - \phi_\alpha)    \,.
\end{equation}
By using the saddle point method we have,
\begin{equation}
	m\phi_m = \min_{\phi} \left[ m\phi + \Sigma(\phi) \right] = m\phi^\star + \Sigma(\phi^\star) \,,
\end{equation}
where $\phi^\star$ is found by $\left. \frac{d \Sigma}{d\phi} \right|_{\phi^\star} = -m$. 
By a Legendre transform we have,
\begin{subequations}
	\begin{align}
		\phi^\star(m) &= \frac{\partial (m \phi_m)}{\partial m}\,,\\
		\Sigma(m) &= - m^2 \frac{\partial \phi_m}{\partial m}    \,.
	\end{align}
\end{subequations}
If $\Sigma(m)<0$ for some $m$ this signals that the solution we found is nonphysical. This usually happens if $m$ is large enough. We show an example of this in~\cref{fig:RSComplexity}, in the case of the square wave activation with $\delta = 1.5$. 

\section{Reinforced Approximate Message Passing}
\label{app:AMPreinforcementApp}

The reinforced approximate message passing algorithm (RAMP) is a message passing heuristic algorithm for sampling solutions in CSPs. In the perceptron case, given a dataset $\mathcal{D}=\{(\boldsymbol{x}^{\mu},y^{\mu})\}$, with $\mu=1,\dots,P$,
generated according to either the storage or the teacher student ensemble, we would like to find a $\boldsymbol{w}\in\{\pm 1\}^N$ such that the constraints
\begin{equation}
	\label{eq:CollisionConstraint}
	y^{\mu}=\varphi\left(\frac{1}{\sqrt{N}}\boldsymbol{w}\cdot\boldsymbol{x}^{\mu}\right),\quad \mu\in[P] \,,
\end{equation}
are satisfied. In order to write down RAMP, let us define for each pattern $\mu=1,\dots,P$ a ``latent field'' $g_{\mu}$, and for each site $i=1,\dots,N$ a ``magnetization'' $m_i$. In the following, we use the convention that Greek indices run from $1,\dots, P$, while Latin indices form $1,\dots,N$. RAMP update iteratively the $m_i$'s and the $g_{\mu}$'s, starting from a random initialization $m_i^0$ and $g_{\mu}^0$. The magnetizations and latent fields at time $t$, respectively $m_i^t$ and $g_{\mu}^t$, are given by: 
\begin{equation}
	\label{eq:AMP1}
	V_{\mu}^{t}=\sum_i(x^{\mu}_i)^2\,\Big(1-\big(m_i^{t-1}\big)^2\Big)\,,
\end{equation}
\begin{equation}
	\label{eq:AMP2}
	\omega_{\mu}^{t}=\sum_ix^{\mu}_i\,m_i^{t-1}-V_{\mu}^{t}\,g_{\mu}^{t-1}\,,
\end{equation}
\begin{equation}
	\label{eq:AMP3}
	g_{\mu}^{t}=f(\omega_{\mu}^{t},V_{\mu}^{t},y_{\mu})\,,\quad \dot{g}_{\mu}^{t}=\partial_{\omega} f(\omega_{\mu}^{t},V_{\mu}^{t},y_{\mu})\,,
\end{equation}
\begin{equation}
	\label{eq:AMP4}
	\Sigma_i^{t}=\left[-\sum_{\mu}(x^{\mu}_i)^2\,\dot{g}_{\mu}^{t}\right]^{-1}\,,
\end{equation}
\begin{equation}
	\label{eq:AMP5}
	h_i^{t}=m_i^{t-1}+\Sigma_i^{t}\,\sum_{\mu}x^{\mu}_i\,g_{\mu}^{t}+r\,t\,\tanh^{-1}(m_i^{t-1}) \,,
\end{equation}
\begin{equation}
	\label{eq:AMP6}
	m_i^{t}=\tanh{(h_i^{t})},
\end{equation}
where the $\omega_{\mu}$, the $V_{\mu}$, the $\dot{g}_{\mu}$, $\Sigma_i$, $h_i$ are auxiliary variables defined in terms of the magnetizations and the latent fields. The function $f(\omega,V,y)$ (and its derivative $\partial_{\omega}f(\omega,V,y)$) is defined by the activation $\varphi(\bullet)$ according to the following formula:
\begin{equation}
	f(\omega,V,y)=\partial_{\omega}\log{\int dz\, \mathrm{e}^{-\frac{(z-\omega)^2}{2V}}\Theta\big(y\,\varphi(z)\big)}\,.
\end{equation}
For the SWP one has:
\begin{equation}
	\begin{split}
		f(\omega,y,V)&=\partial_{\omega}\log{\left(1-y\frac{4}{\pi}\sum_{n=0}^{\infty}\frac{\mathrm{e}^{-\pi^2\,\frac{V}{\delta^2}\frac{(2n+1)^2}{2}}}{2n+1}\sin{\left(\frac{\pi}{\delta}(2n+1)\,\omega\right)}\right)}=\\&= -y\frac{4}{\delta}\frac{\sum_{n=0}^{\infty}\mathrm{e}^{-\pi^2\,\frac{V}{\delta^2}\frac{(2n+1)^2}{2}}\cos{\left(\frac{\pi}{\delta}(2n+1)\,\omega\right)}}{1-y\frac{4}{\pi}\sum_{n=0}^{\infty}\frac{\mathrm{e}^{-\pi^2\,\frac{V}{\delta^2}\frac{(2n+1)^2}{2}}}{2n+1}\sin{\left(\frac{\pi}{\delta}(2n+1)\,\omega\right)}}\,.
	\end{split}
\end{equation}
\begin{equation}
	\partial_{\omega}f(\omega,y,V)=-f^2(\omega,y,V)+y\frac{4\pi}{\delta^2}\frac{\sum_{n=0}^{\infty}\mathrm{e}^{-\pi^2\,\frac{V}{\delta^2}\frac{(2n+1)^2}{2}}(2n+1)\sin{\left(\frac{\pi}{\delta}(2n+1)\,\omega\right)}}{1-y\frac{4}{\pi}\sum_{n=0}^{\infty}\frac{\mathrm{e}^{-\pi^2\,\frac{V}{\delta^2}\frac{(2n+1)^2}{2}}}{2n+1}\sin{\left(\frac{\pi}{\delta}(2n+1)\,\omega\right)}}\,.
\end{equation}
We call ``iteration'' an update of all magnetizations (and therefore of all the latent fields). The parameter $r$ in \eqref{eq:AMP5} is called reinforcement rate. Setting $r=0$, RAMP becomes equivalent to the approximate message passing (AMP) algorithm, which is used to estimate the local marginals of variables in constraint-satisfaction problems \cite{mezard2009information}. In the teacher-student case, for $\alpha\geq\alpha_r$, the local magnetizations polarize in the direction of the teacher. An example is given in \cref{fig:amp_success_rate,fig:amp_spinodal_transition}. This is the only situation where AMP provides a solver for \cref{eq:CollisionConstraint}. In other cases, AMP returns non-integer magnetization that do not define a vertex of the hypercube. Setting $r>0$, one introduces in the AMP iteration a term that tends to polarize the magnetizations, providing a heuristic solver for the problem  \cite{chavas2005survey,braunstein2006learning}. In this case, equations (\ref{eq:AMP1})-(\ref{eq:AMP6}) are iterated until the binary vector $\tilde{\boldsymbol{x}}^t$, with elements $\tilde{x}^t_i=\text{sgn}(m_i^t)$, satisfies constraint \ref{eq:CollisionConstraint}, or $t$ becomes larger that a maximum time $T_{max}$, that is a parameter of the algorithm. There are different choices of the reinforcement term suggested in the literature, which all lead to qualitatively similar results. The behavior of the solver depends on the choice of $r$.  The performance of the solver, which is determined by the largest value of $\alpha$ at which the solver is able to find solutions in polynomial time, improves by reducing $r$. However, it is also found that the number of iterations required to find a solution scales like $1/r$. Therefore, it is convenient to define the optimum reinforcement rate $r_{opt}$, which is the largest (therefore the cheapest) reinforcement rate that allows finding a solution \cite{braunstein2006learning,Baldassi2015}. In \cref{fig:simulazioniampreinforcement} we show $T_{sol}$, which is the average number of iterations required to find a solution when the reinforcement rate is chosen as the optimal one, as a function of $\alpha$ for different values of $\delta$.  

\end{document}